\documentclass{article}
\usepackage{jfrExamplee}
\usepackage{graphicx}
\usepackage{apalike}
\usepackage{setspace}
\usepackage{amsmath}
\usepackage{multirow}
\usepackage{lineno}
\usepackage{epstopdf}
\usepackage{booktabs} 
\usepackage{tabularx} 
\usepackage [english]{babel}
\usepackage [autostyle, english = american]{csquotes}
\MakeOuterQuote{"}

\title{All Weather Perception: Joint Data Association, Tracking, and Classification for Autonomous Ground Vehicles}

\author{
Peter Radecki\thanks{emails: \texttt{ppr27@cornell.edu, mc288@cornell.edu, kmatzen@cs.cornell.edu}}\ \ and Mark Campbell \\
Sibley School of Mechanical and Aerospace Engineering\\
\And
Kevin Matzen\\
Computer Science Department\\
\AND
\textnormal{Cornell University}\\
Ithaca, NY 14853 \\
}

\begin{document}

\maketitle

\begin{abstract}
A novel probabilistic perception algorithm is presented as a real-time joint solution to data association, object tracking, and object classification for an autonomous ground vehicle in all-weather conditions. The presented algorithm extends a Rao-Blackwellized Particle Filter originally built with a particle filter for data association and a Kalman filter for multi-object tracking \cite{RBPF-Skynet} to now also include multiple model tracking for classification. Additionally a state-of-the-art vision detection algorithm that includes heading information for autonomous ground vehicle (AGV) applications was implemented. Cornell's AGV from the DARPA Urban Challenge was upgraded and used to experimentally examine if and how state-of-the-art vision algorithms can complement or replace lidar and radar sensors. Sensor and algorithm performance in adverse weather and lighting conditions is tested. Experimental evaluation demonstrates robust all-weather data association, tracking, and classification where camera, lidar, and radar sensors complement each other inside the joint probabilistic perception algorithm.

\end{abstract}

\section{Introduction}
\label{intro}
The past decade has seen rapid advancement in autonomous ground vehicles (AGV) from academic research projects: five vehicles successfully completed the second DARPA Grand Challenge and six completed the DARPA Urban Challenge; and industry vehicle research: numerous companies from traditional automotive sectors such as General Motors and Ford, military sectors such as Oshkosh Trucks, and information technology sectors such as Google, safely driving hundreds of thousands of miles with autonomous vehicles on public roads; to public policy advancement: Nevada became the first state to license autonomous vehicles and semi-trucks. Both the NHTSA \cite{NHTSA-guidelines} and SAE \cite{SAE-guidelines} have published roadmaps for the future development of autonomous vehicles and have posed the Holy Grail of "Level 4 or 5" as full autonomy in any environment and situation. One might exclude unique situations such as a wild African Safari through the unmarked bush, but commercial autonomy fit for public use is generally understood to include standard driving on all modern road systems. A key shortcoming of current work is autonomous driving in all weather conditions. This is precisely where the limits of current advanced driver assistance systems (ADAS) as well as AGV development and testing are.

At the 2014 Future Automobile Technology Competition in South Korea, rain fell the morning of the second day of testing \cite{KAIST-weather}. The result? Two autonomous vehicles that navigated the course successfully on the first day crashed on the second---predominantly a result of perception system failures solely due to a wet road and humid conditions. The message was clear: autonomous vehicles are not ready for public adoption until they have been validated and tested to work in all-weather scenarios. While it is true that prudence can cause even the most talented drivers to avoid certain weather conditions, realistic public acceptance of fully automated driving capabilities requires robustness to reasonable and common weather and lighting phenomenon such as snow, rain, fog, and night conditions. The aim of this paper is two-fold. First, a set of data logs are collected in varying weather and lighting conditions to evaluate perception algorithms. Second, perception robustness in all-weather conditions is improved by a joint Bayesian solution to association, tracking, and classification that includes state-of-the-art vision algorithms in addition to lidar and radar sensors.

Since the DARPA Challenges, many advances have occurred in computer vision, lidar point cloud segmentation, and vehicle embedded computing---all of which have direct application in autonomous vehicles. The field of vision-based object detection and localization has experienced several notable innovations such as the development of (1) good low-level feature descriptors that capture local shape, but remain invariant to local photometric and geometric changes \cite{HOG-detections}, (2) models that capture larger scale deformations not captured by the low-level features themselves and methods for the discriminative training of these models \cite{DPM-TPAMI}, and (3) deep learning methods that learn a rich hierarchy of low-level to high-level features from data with little to no manual engineering of the model structure \cite{deep-learning}, \cite{Girshick-CVPR-2014}, \cite{Girshick-ICCV-2015}, \cite{NIPS-2015}, and \cite{He-arxiv-2015}. The field of deep learning and deformable parts models \cite{deep-learning-tutorial} and \cite{Felzenszwalb-DPM} has advanced from hand-designed feature detection methods to learn hierarchies of features in an unsupervised manner directly from data, vastly improving algorithm detection and classification accuracy. Lidar point cloud processing methods have improved to enable accurate segmentation and classification of high resolution 3d point clouds in real-time \cite{Douillard-ICRA2011} and \cite{Korchev-RTLidar}. Aligning point-clouds with iterated closest point methods has been shown to improve tracking performance of obstacles' absolute ground speed, an inherently noisy parameter when estimated as a derivative of position in a parametric filter \cite{Held-ICRA2013}. The entire field of GPU processing for parallelizable computation was developed with the introduction of NVIDIA's proprietary CUDA platform and later support by AMD and Nvidia for the Open Computing Language (OpenCL), enabling real-time computation of that which was previously relegated to a cluster or server farm. Further advancements have demonstrated real-time performance of DPM vision detection methods \cite{Held-ICRA2012} and \cite{DPM-30Hz}, and hardware advances in embedded computing have shown deep learning classification running on rugged mobile platforms \cite{DrivePX}.

With all the recent advances, there have also been some patterns highlighting the current opportunities for future development. With a few exceptions, the vast majority of published studies to date have demonstrated the performance of autonomous vehicles in optimal weather conditions, such as sunny daytime. For example, many well published autonomous driving research efforts have focused their testing in the sunny, fair-weather areas of California and Nevada. This motivates the need to understand how sensors and perception algorithms handle changing weather conditions.

\begin{figure} [h]
    \centering
    \includegraphics[width=6.0in]{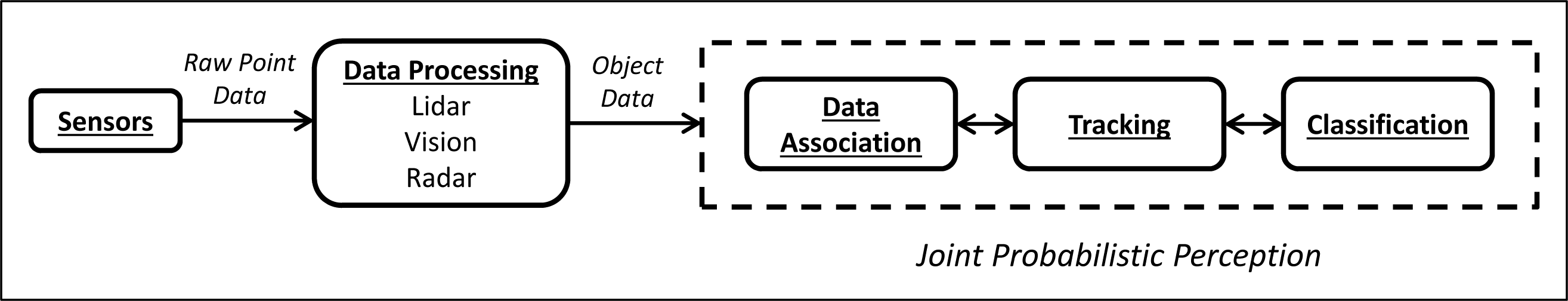}
    \caption{Typical AGV perception pipeline.}
    \label{fig:perception-pipeline}
\end{figure}

Figure~\ref{fig:perception-pipeline} shows the major components commonly found in AGV systems, known generally as segmentation and clustering (processing raw sensor analog or digital data into obstacle-level meta-measurements), data association (determining which  measurement came from which static or dynamic environmental obstacle), tracking (estimating the obstacle's state, position, velocity, etc.), and classification (distinguishing cars, people, buildings, etc.). \cite{Teichman-ARSO2011} stated that full joint solutions to this perception problem are intractable to formulate or compute. Many advanced techniques recently published have evaluated performance of some of these components in isolation from others in the overall perception pipeline \cite{Teichman-ICRA2013}, \cite{Ilg-ICRA2014}, \cite{Held-RSS-14}. Examples include: improved tracking and classification algorithms that ignored any segmentation or data association errors \cite{Teichman-ARSO2011}, performed evaluations on very limited types of scenarios such as tracking a large number of stationary cars or a small number of dynamic objects \cite{Held-ICRA2013}, simply lacked access to large public-domain accurately labeled urban data sets available to quantitatively evaluate performance \cite{Korchev-RTLidar}, or handled classification separately after combining data association and tracking \cite{RBPF-Skynet}. These aforementioned studies have plenty of merit, but the lack of a joint solution is both a concern in practical solutions for an AGV and also an opportunity for development. This paper presents a full joint solution implemented in real-time and tested in both real-world urban environments and repeatable staged scenarios for qualitative and quantitative evaluation across different weather conditions.

A corresponding philosophical question arises in robotics literature \cite{Probabilistic-Robotics} regarding the importance of actually modeling the dynamics of all possible object classifications if the inputs are truly unknown. In a vehicle, changes in speed and direction are almost entirely due to driver inputs limited by vehicle performance, which in practice can negate the ability of a kinematics classifier to differentiate object types. For example, a conservative driver may drive a Corvette with the same slow acceleration as an 18-wheeler semi-truck. Despite the difficulty, classification is innate to human driving and perception; the moment before an accident one may attempt to quickly maneuver. Understanding if the object on the side of the road is a human, a shrub, or a boulder is critical to that crash avoidance decision. When driving, a human classifies objects based on their appearance and actions. A camera may detect if an object is a person or a car, but by their motion a person can infer if extra caution must be exercised around the object, such as children playing in the street or a distracted driver. There is utility in Bayesian inference of classification combining sensor detected information and kinematics tracking information. One example approach might be to utilize a bank of filters to detect normal driving versus erratic driving; classification could be passed to a safety-conscious system to leave a larger berth around erratic vehicles. There are some computational concerns which arise, but the number of unique kinematic classes of dynamic roadway objects is small and in practice this inherently helps limit the maximum computational overhead required to support classification.

With the rapid development in the field, the architecture for a proto-typical autonomous ground vehicle is not yet set in terms of sensing, computation hardware, or programming interfaces. Google, for example, heavily utilizes background subtraction for lidar sensor processing, subtracting sensor returns from a pre-built 3D static environmental map to identify moving obstacles \cite{Urmson-2011}. Of any company, Google may have the best infrastructure to 3d map out every public road in the United States, but one can readily point out operational limitations of such a system to exclude, off-road excursions, private driveways/roads, road-construction re-routing, theater of war, and operation after a natural disaster---a time when one may be most dependent on vehicle mobility. This motivates an interesting question: is \textit{a priori} mapping required or even necessary? Other companies have focused on building systems around different combinations of radar, camera, infrared, lidar, and ultrasonic sensors \cite{Mercedes-S-Class} and \cite{agv-wired}.

Given the current variances in AGV sensing and architecture, and in an effort to understand the challenges adverse weather and lighting scenarios pose, this study examines how the different sensing modalities of radar, lidar, and camera perform in diverse environmental conditions including snow, rain, fog, and nighttime. A particular focus is given to understanding how state of the art vision processing compares to lidar---the predominant sensor of most successful DUC teams. In adverse weather, assumptions about other layers of the perception pipeline can no longer be guaranteed. Furthermore, different layers of the perception pipeline potentially can aid one another---object classification may inform the dynamics of the object to better track it, or existing object tracks may aid data association for region-of-interest image detections. To highlight an example of potential capabilities joint perception solutions provide, a demonstration of object classification based on multiple model Kalman filter tracking is presented for objects in an urban environment. Such multiple model tracking has been shown beneficial for tracking airplanes in turning and straight flight \cite{mm-kinematics-classification} but is novel for classification of terrestrial objects in urban vehicular environments. Kinematics-based classification might be beneficial in precipitation, as camera detections or lidar returns become obscured, or when tracking solely with radar. The authors believe that one of the keys to handling adverse environmental operating conditions is a full Bayesian probabilistic joint perception system, thereby minimizing the number of brittle ad hoc design choices which tend to fail under uncommon untested weather scenarios. This paper utilizes Skynet, Cornell's autonomous 2007 Chevrolet Tahoe from the DUC, and builds upon \cite{RBPF-Skynet} to extend joint data association and tracking to include classification; relaxations allowing computational feasibility come from a Rao-Blackwellized Particle Filter (RBPF), multiple hypothesis modeling, and carefully managing measurements in forward-pass parametric filters.

In summary this paper makes the following contributions:
\begin{itemize}
\item Demonstrates object classification in an urban environment based on multiple model tracking.
\item Demonstrates a real-time joint probabilistic method to solve data association, tracking, and classification for an AGV roadway environment.
\item Examines if and how state-of-the-art vision algorithms can compliment or replace lidar and radar sensors.
\item Investigates sensor and perception algorithm performance in adverse weather and lighting conditions.
\end{itemize}

\section{Joint Probabilistic Formulation}
Before deriving a full Bayesian formulation for joint data association, tracking, and classification, a brief example is given to demonstrate how measurements fed into a tracker can be used to correctly classify the object solely based on dynamics without any sensor-specific meta-information on the object's shape, size, color, or type. Kinematics-based classification methods which match an object's dynamics model with measured data points typically require access to measurement update residuals (innovations), and covariances from a Kalman Filter (KF) or Particle Filter (PF). By building upon the brief classification example, a full Bayesian formulation is then developed that extends the combined data association and tracking RBPF from \cite{RBPF-Skynet}.

\subsection{Joint Classification and Tracking -- Derivation and Example}
In contemporary literature, direct object classification typically focuses on image processing techniques such as feature extraction or constructing a lidar point cloud and comparing against pre-classified 3d models. Similar work has been done by \cite{Montiero-IROS2006} to combine a KF with extra sensor information such as imagery data to infer object classification. Alternative approaches known as boosting methods have combined banks of weak classifiers to infer object classification; AdaBoost is one of the most popular \cite{Adaboost}.

Reliable inference of classification is accomplished here by extending a standard Kalman filter tracker given noisy position information of a target, which could be collected from any standard AGV sensor such as radar, camera, or lidar. Separate classifiers are designed for cars, pedestrians, cyclists, and buses--four of the most common moving objects in an urban environment.

\begin{figure} [h]
    \centering
    \includegraphics[height=1.0in]{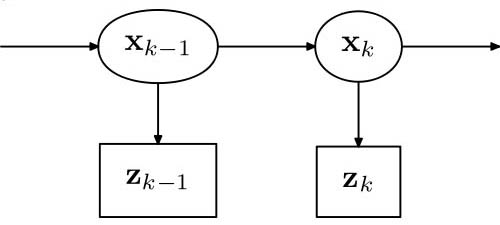}
    \caption{Kalman Filter as Hidden Markov Model.}
    \label{fig:kf-hmm}
\end{figure}

Kalman Filter derivations depend on the inherent uncertainty in the system dynamics and measurements. By assuming Gaussian distribution uncertainty, the modeled system can be viewed as a mixture or compilation of Gaussian distributions.  In a probabilistic graphical models framework, a standard KF is a hidden Markov Chain that has observed noisy measurements where the chain describes the evolution of the dynamic system through time.  This chain is illustrated in Figure~\ref{fig:kf-hmm}, where $\mathbf{x}$ is the hidden system state and $\mathbf{z}$ is the observed measurement.  Implementing a KF requires knowledge of the process, dynamics, process noise, and measurement noise.  Combining this information along with an estimate of the state initial condition, the KF operates online, updating its estimate or inference of the system state at each time step $k$, calculating $p(\mathbf{x}_k | \mathbf{z}_k)$.

\begin{figure} [h]
    \centering
    \includegraphics[height=1.5in]{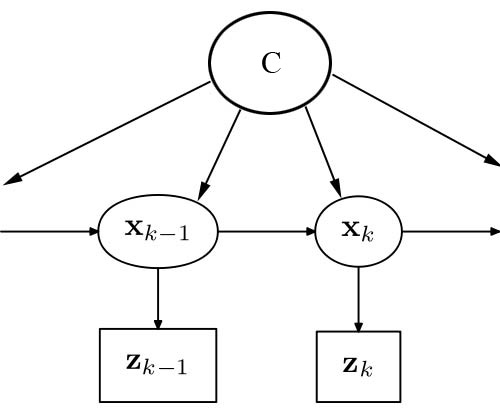}
    \caption{HMM dependent on model classification.}
    \label{fig:kf-class-hmm}
\end{figure}

Measurements are assumed to be received from a specified unknown object whose classification exists within a set containing uniquely modeled dynamic processes. As shown in Figure~\ref{fig:kf-class-hmm}, the object classification $\mathbf{C}$ is inferred based on the correlation between the dynamic model and the measurements as $p(\mathbf{C}|\mathbf{z}_k)$. Developing a KF requires proper tuning of noise parameters in order to best match the model with the physical system.  Innovation test statistics are typically used to validate this matching. The example here uses innovation statistics in a batch methodology to associate the correct dynamic model with the measurement data. Reliably inferring the classification requires accurate and computationally simple models of the dynamic processes.

\begin{figure} [h]
    \centering
    \includegraphics[height=1.2in]{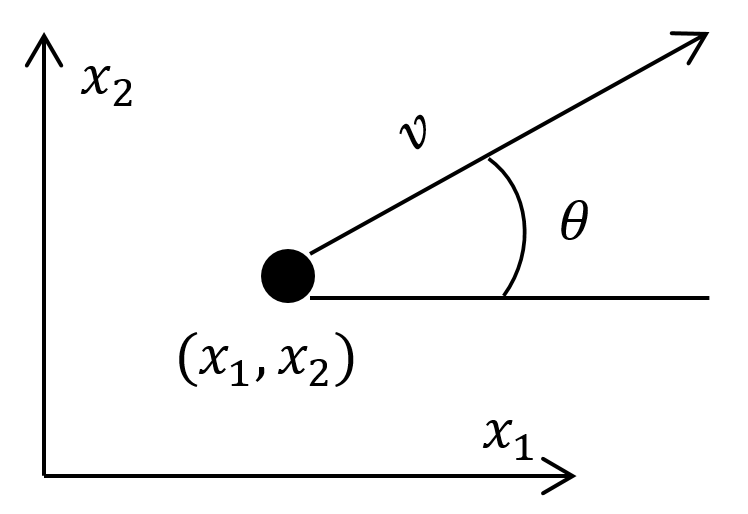}
    \caption{Coordinate system used in differential equations.}
    \label{fig:coord-sys}
\end{figure}

Pedestrians can walk in any direction while cars, buses, and cyclists are subject to non-holonomic constraints of rolling wheels. Inputs are unmodeled so the process noise terms must account for all changes of direction and speed. The dynamics model of the person is assumed to have Gaussian process noise equal in $x_1$ and $x_2$ directions given by the following differential equations:
\begin{equation}
\begin{aligned}
\ddot{x}_1 &= e_{x_1} \\
\ddot{x}_2 &= e_{x_2}
\end{aligned}
\end{equation}
 where $e_{x_1}$, and $e_{x_2}$ correspond to the acceleration process noise in the East and North Cartesian directions, respectively, similar to that presented in \cite{KF-Radar}. Using the four state representation, $\mathbf{x} = \begin{bmatrix} x_1 & x_2 & \dot{x}_1 & \dot{x}_2\end{bmatrix}^\top$, a linear KF is used for inference over the person model. The differential equations for the wheeled objects are:
\begin{equation}
\begin{aligned}
\dot{x}_1 &= v \cos(\theta) \\
\dot{x}_2 &= v \sin(\theta) \\
\dot{v} &= e_v \\
\dot{\theta} &= e_{\theta}
\end{aligned}
\end{equation}
with state $\mathbf{x} = \begin{bmatrix} x_1 & x_2 & v & \theta\end{bmatrix}^\top$,  where $e_{x_v}$, and $e_{x_\theta}$ correspond to the acceleration process noise in the velocity and heading directions, respectively. An Extended Kalman Filter is used for inference over the car, bus, and cyclist models. Figure~\ref{fig:coord-sys} illustrates the variables used in the coordinate system.

Using the innovation and innovation covariance, which are based on the state estimate and most recent measurement, a statistic $d^2$ is defined as
\begin{equation}
\begin{aligned}
d_k^2 = \tilde{y}_k^\top S_k^{-1} \tilde{y}_k
\label{eq:d-squared}
\end{aligned}
\end{equation}
where $\tilde{y}$ is the measurement innovation or residual, $S$ is the innovation covariance, and $d^2$ is a chi-squared, $\chi^2$, distributed variable with a number of degrees of freedom equal to the length of the measurement vector (2 for this scenario). The $d^2$ statistic, called the Normalized Innovation Squared (NIS), relates the probability of correlation between the measurement and the model as given by \cite{Bar-Shalom}. A bad measurement or a bad model yields a poor $d^2$ statistic. Engineers use this $d^2$ variable online to gate erroneous measurements within a selectable confidence interval, such as a $2\sigma$, 95\% bound, by comparing $d^2$ against the $\chi^2$ values. To evaluate the likelihood of the model, instead of the likelihood of an individual measurement, the test statistic is averaged over $k$ Kalman filter iterations to get $\bar{\epsilon}_k$ and compared to the associated chi-squared probability with $2\times k$ degrees of freedom:
\begin{equation}
\begin{aligned}
\bar{\epsilon}_k = \frac{1}{k} \sum_{i=1}^{k} d_i^2
\end{aligned}
\end{equation}
Comparing values of $\bar{\epsilon}_k$ for different models and picking the one with the highest probability yields a simple, reliable approach to infer object classification using the given measurement track. Note that while the presented formulation assumes a batch processing methodology, likelihood estimation is used in the later developed fully joint solution to estimate object classification in real-time.

\subsection{Joint Data Association, Tracking, and Classification}
The following derives a joint perception solution for data association, tracking, and classification. The perception platform of an autonomous vehicle must take in measurements $Z$ from a set of on-board sensors to estimate its local environment which is made up of objects $O$ with unique dynamics which have a classification $C$. The discrete variable $A$ assigns each measurement $Z$ to the object $O$ from which it originated. In \cite{RBPF-Skynet} a joint solution for measurement association and object tracking is presented as a Rao-Blackwellized Particle Filter which solves $p(A_k,O_k|Z_k)$, where capital letters with $k$ subscripts represent the set's history until time $k$. The sought after joint solution is that of the distribution
\begin{equation}
p(A_k, O_k, C_k|Z_k).
\end{equation}
Some general filtering methods such as the Gaussian Mixture Probability Hypothesis Density Filter \cite{GMPHD-Filter} exist which could estimate joint densities over different variables, but in general, no closed form solutions exist, and computational requirements for an exact filter would grow exponentially through time and would be impractical. However, there are some unique aspects of the perception problem that can be used to intelligently split the problem and make it feasible. First, the system is hybrid: object states are continuous while measurement assignments are discrete. Object classifications are discrete and typically time invariant.\footnote{A note on classification time invariance: people may exit a car, but the car has not changed classification, rather a new  object has entered the observable scene. However, some classifications could be time-varying, for example the classification of an object as static (stationary) or dynamic (moving)---a parked car may start driving.} The number of possible data assignments and time-variant classifications grows exponentially with time, number of objects, and  number of measurements, rendering exact probabilistic reasoning impossible even for simple scenarios.

Miller showed how the infeasible problem can be made feasible by using factorization and sampling techniques \cite{RBPF-Skynet}. First, the joint solution is factored exactly as
\begin{equation}
p(A_k, O_k, C_k|Z_k) = p(A_k|Z_k) p(O_k|Z_k, A_k) p(C_k|Z_k, A_k, O_k).
\end{equation}
This factorization provides an intuitive decoupling of the problem into discrete assignment $p(A_k|Z_k)$, object tracking $p(O_k|Z_k, A_k)$, and classification $p(C_k|Z_k, A_k, O_k)$. Decoupling helps enable solutions to tractable sub-problems, but does not fix the computational intractability due to the problem's exponential growth through time. By sampling the discrete data assignment density $p(A_k|Z_k)$, Monte Carlo likelihood-weighted techniques can simplify the computational complexity. Furthermore, the dynamics' differential equations can be modeled in state-space giving them a Markov property that only the current state need be saved for a given object. The continuous dynamics can then be readily estimated by parametric filtering such as the Kalman Filter. By keeping only the current object state, the prior data assignment history also is not required to be kept. In \cite{GMPHD-Filter} this same splitting was implemented via a Rao-Blackwellized Particle Filter (RBPF). For the sake of brevity, a full derivation of particle likelihood formulations including birth and death likelihoods and resampling procedures has been omitted; \cite{GMPHD-Filter} and \cite{RBPF-Skynet} both present thorough summaries. This paper extends these formulations by specifically adding the classification $p(C_k|Z_k, A_k, O_k)$ term.

In summary, in the RBPF formulation, each particle represents one hypothesis of the scene that includes all measurements and object states through time. Measurement assignment likelihoods are drawn across all objects in a given particle. The states for each object in a particle are predicted forward in time and updated for the assigned measurement. Different particles thus represent different measurement assignment histories as well as their corresponding updates. Object births and deaths are particle specific, and particles need not have the same number of objects.

As presented in the introductory classification example, object dynamics are dependent on classification. If an object's true classification is known \textit{a priori}, the correct dynamical model is used in the tracker. In practice, model type is not known \textit{a priori}, so the formulation here tracks objects using a bank of filters spanning the set of possible dynamics. The probability of classification can then be written as
\begin{equation}
\begin{aligned}
p(C|Z_k, A_k, O_k) &= \frac{p(C,Z_k| A_k, O_k)}{p(Z_k|A_k, O_k)} \\
 &= \frac{p(C,z_k,Z_{k-1}|A_k, O_k)}{p(z_k,Z_{k-1}|A_k, O_k)}\\
 &= \frac{p(C,z_k|Z_{k-1},A_k, O_k)p(Z_{k-1}|A_k, O_k)} {\sum\limits_{j=0}^{n_c} p(C=j,z_k,Z_{k-1}|A_k, O_k)} \\
 &= \frac{p(z_k|C,Z_{k-1},A_k, O_k)p(C|Z_{k-1},A_k, O_k)p(Z_{k-1}|A_k, O_k)} {\sum\limits_{j=0}^{n_c} p(z_k|C=j,Z_{k-1},A_k, O_k)p(C=j|Z_{k-1},A_k, O_k)p(Z_{k-1}|A_k, O_k)} \\
 &= \frac{p(z_k|C,A_k, O_k)p(C|Z_{k-1},A_{k-1}, O_{k-1})} {\sum\limits_{j=0}^{n_c} p(z_k|C=j,A_k, O_k)p(C=j|Z_{k-1},A_{k-1}, O_{k-1}))} \label{eq:prob_class}
\end{aligned}
\end{equation}
where $n_c$ is the number of possible classification categories. The formulation given in (\ref{eq:prob_class}) is for classification that is time invariant. Time varying classification can easily be developed by extending (\ref{eq:prob_class}) as the following recursion:
\begin{equation}
\begin{aligned}
p(c_k|z_k, a_k, o_k)  &= \frac{p(z_k|c_{k-1},a_k, o_k)p(c_{k-1}|z_{k-1},a_{k-1}, o_{k-1})} {\sum\limits_{j=0}^1 p(z_k|c_{k-1}=j,a_k, o_k)p(c_{k-1}=j|z_{k-1},a_{k-1}, o_{k-1}))}
\end{aligned}
\end{equation}
In the case of time-varying classification, a process noise term forgetting factor could be included in the likelihood formulation in order to allow the estimate to vary based on a time-constant.

Particles within the RBPF are drawn according to a proposal density $q(A_k|Z_k)$ selected for its efficient sampling algorithms and similarity to $p(A_k|Z_k)$. Particles have a weight and diversity such that they span and represent $p(A_k|Z_k)$ as follows
\begin{equation}
\begin{aligned}
p(A_k|Z_k) &\approx \sum\limits_i w_k^i \delta(A-A_k^i) \\
w_k^i &= \frac{p(A_k^i|Z_k)}{q(A_k^i|Z_k)} \\
\sum\limits_i w_k^i &=1 \label{eq:weight_def}
\end{aligned}
\end{equation}
where $w_k^i$ is the likelihood weight of the $i$th particle $A_k^i$ at time index $k$, and $\delta( \cdot )$ is the Kronecker delta function for discrete assignment. Given the factorization of $q(A_k|Z_k)$ as follows
\begin{equation}
\begin{aligned}
q(A_k|Z_k) = q(a_k|Z_k,A_{k-1})q(A_{k-1}|Z_{k-1})
\end{aligned}
\end{equation}
the likelihood weight $w_k^i$ can be expressed recursively as
\begin{equation}
\begin{aligned}
w_k^i \propto \frac{\sum\limits_{j=1}^{n_c}p(z_k|Z_{k-1},A_k^i,c_{k-1}^i=j)p(a_k^i|Z_{k},A_{k-1}^i)}{q(a_k^i|Z_{k},A_{k-1}^i)}w_{k-1}^i
\end{aligned}
\end{equation}
where the symbol $\propto$ indicates weights must be renormalized after update to maintain unity summation from equation (\ref{eq:weight_def}).

Within the RBPF framework, objects are initialized with a bank of $n_c$ KFs, one KF for each possible unique object classification. The question naturally arises of how to calculate the association likelihood for a given object in particle $i$ against a bank of KF model classifications. Given the normalized classification probability, as is the case for a unique, mutually exclusive classification set, the summation of probabilities of all $n_c$ unique possible classifications for a given object is $\sum\limits_{j=1}^{n_c} p(c_k=j|z_k, a_k, o_k) = 1$. For a parametric KF the optimal proposal density can be directly sampled. Thus the likelihoods and normalizing factor terms for sampling the overall particle filter can be given as
\begin{align}
q_{\text{opt}}(a_k^i|Z_k,A_{k-1}^i) &= \sum\limits_{j=1}^{n_c}  \alpha_k^i p(c_{k-1}=j|Z_{k-1}, A_{k-1}^i, O_{k-1}^i)p(z_k|a_k^i,Z_{k-1},A_{k-1}^i,c_{k-1}^i=j) p(a_k^i|Z_{k-1},A_{k-1}^i) \label{eq:prop-density}\\
\alpha_k^i &= \left[\frac{1}{M^i} \sum\limits_{m=1}^{M^i} \sum\limits_{j=1}^{n_c} p(c_{k-1}=j|Z_{k-1}, A_{k-1}^i, O_{k-1}^i) p(z_k|a_{m,k}^i ,Z_{k-1},A_{k-1}^i,c_{k-1}^i=j)\right]^{-1} \label{eq:prop-density-norm}
\end{align}
where $n_c$ is the total number of classifications, $j$ represents the selected classification, $i$ represents the selected particle number, $m$ represents the selected object in a particle, $M^i$ is the total number of tracked objects in the $i$th particle, $q_{\text{opt}}$ is the optimal proposal density from which particles are drawn, and $\alpha_k^i$ is a normalizing constant that depends on the $i$th particle prior to sampling. 

The term $p(a_k^i|Z_{k-1},A_{k-1}^i)$ is the transition model relating \textit{a priori} assignment information $A_{k-1}^i$ to the current measurement. For a generic position or velocity measurement, such as what is obtained from a radar sensor, this transition probability is uniform because previous assignments provide no information about future assignments. However, if one has a camera or point cloud, this probability could relate the previous camera's region of interest to the new measurement, or match the new lidar cluster to an existing built-up point cloud. The probability could even incorporate meta-information such as the color of the car; for example, if tracking a red car, the next camera detection could have its association likelihood modified based on how well the color matches.

The term $p(z_k|a_k^i,Z_{k-1},A_{k-1}^i,c_{k-1}^i=j)$ is the likelihood the measurement originated for a specific tracked obstacle with the $j$th classification in the $i$th particle. Each obstacle in the particle has a bank of $n_c$ associated KFs, each estimating (also known as tracking) the object state according to the classification specific dynamics. For a single classification, a single KF is used and the likelihood is simply calculated from the normalized innovation, similar in concept to the normalized innovation used to classify tracks in the introductory example in Equation (\ref{eq:d-squared}). With a bank of $n_c$ KFs, the normalized innovation must be calculated for each KF in the bank. The likelihood value is then weighted by the respective classification probability $p(c_{k-1}=j|Z_{k-1}, A_{k-1}^i, O_{k-1}^i)$ and summed across all $n_c$ KF tracks in that bank. Intuitively a hard decision on an object classification is naturally made by setting the classification $C=j$ and $n_c=1$ while not adversely affect the Bayesian formulation.

A high-level description of the joint data assignment-tracking-classification algorithm is given as follows.
\begin{enumerate}
	\item Draw an initial set of particles $A_0^i\, \forall\; i\; \exists\, [1,N]$.
	\item Predict all obstacles in each particle forward in time to the next measurement $z_k$ to yield a parametric representation of $p(O_k|Z_{k-1},A_{k-1}^i)$.
	\item Randomly sample a new set of data assignments for $z_k$ from the optimal proposal density according to Equation (\ref{eq:prop-density}) and the given sampling procedure.
	\item For each object in each particle, update the bank of parametric-tracking filters to yield $p(O_k|Z_k,A_k^i)\, \forall\, C$ and update the classification $p(c_k|Z_k,A_k,O_k)$.
	\item Update particle weights according to $w_k^i \propto w_{k-1}^i / \alpha_k^i$ and (\ref{eq:prop-density-norm}), and renormalize the weights to sum to unity.
	\item Resample particles to keep the filter well conditioned, if necessary.
\end{enumerate}

Finer grained classification requires sensor-specific data processing and techniques. For example one could build a 3d colorized point cloud of the object and compare it to a library. The formulation included here natively supports sensor-specific output of object classification along with weak classifiers typically used in a boosting framework \cite{Adaboost}. Boosted weak classifiers and sensor-specific classification information could be incorporated into the likelihood formulation of $p(z_k|c_{k-1},a_k, o_k)$. Experimental studies in the results section of this paper demonstrate classification performance informed solely with lidar clustering and vision processing of object classification.

\subsection{Classification with Multiple Hypothesis}
Vision-based car detections provide vehicle heading in addition to locating the vehicle's bounding box within the scene \cite{nyc3dcars}. Typically the detector can correctly extract the major-axis line of the vehicle, but it often confuses the front and back of the car along that line. This causes the angular heading measurement to have a bi-modal distribution with a main peak along the forward direction of the car and a secondary smaller peak in the 180 degree reverse direction. Handling this distribution as a single Gaussian would require a very large distribution over the angle range of $-\pi$ to $\pi$ radians. Furthermore, the propagation of the position dynamics is coupled to both the heading and ground speed, so a large covariance in the heading estimate causes the filter's overall position and velocity estimates to become uncertain and degrades future measurement association performance.

Multiple options exist for handling multi-modal distributions, such as Gaussian Sum Filters \cite{GSF} or Particle Filters \cite{posterior-pose}. An alternative approach is to add a state variable for vehicle direction, either forward or reverse, and treat the measurement in two parts 1) as the angle to a line and 2) a binary variable corresponding to the forward direction along the heading line. This measurement splitting allows for a simple formulation with minimal additional computation by augmenting the Kalman Filter continuous-state tracker with a Bayesian classifier for the discrete random variable of the vehicle's forward direction. The following formulation allows for tracking vehicles driving forward or reverse. It correctly distinguishes the true vehicle orientation with the only requirement being that the camera (or other sensor) has a weak classifier of vehicle forward direction; that is, the classifier must correctly distinguish forward from reverse in more than 50\% of its heading detections.

Along the central axis of the car, two classifications $H$ for heading exist as the set $h\, \exists\, \{0 = \tt{reversed},$ $1 = \tt{correct}\}$. The classifier $p(H|Z_{k})$ is written as
\begin{equation}
\begin{aligned}
p(H|Z_{k}) &= \frac{p(H,Z_{k})}{p(Z_{k})} \\
 &= \frac{p(H,z_k,Z_{k-1})}{p(z_k,Z_{k-1})}\\
 &= \frac{p(H,z_k|Z_{k-1})p(Z_{k-1})} {\sum\limits_{h=0}^1 p(H=h,z_k,Z_{k-1})} \\
 &= \frac{p(z_k|H,Z_{k-1})p(H|Z_{k-1})p(Z_{k-1})} {\sum\limits_{h=0}^1 p(z_k|H=h,Z_{k-1})p(H=h|Z_{k-1})p(Z_{k-1})} \\
 &= \frac{p(z_k|H)p(H|Z_{k-1})} {\sum\limits_{h=0}^1 p(z_k|H=h)p(H=h|Z_{k-1})}
 \label{eq:heading_class}
\end{aligned}
\end{equation}
The last line can be read as the $k^{th}$ detection likelihood times the prior divided by a normalizing factor. As long as the camera's classification detection probability for a given detection exceeds 50\%, this sequential estimator correctly classifies the probability of the heading. Additionally, if initialization is based on object location with respect to lanes, the initial prior $p(H|Z_{0})$ could be nonuniform and heavily weighted for a car obeying rules of the road and traveling in the direction of traffic flow. Given a uniform prior, the maximum \emph{a posteriori} of the classifier simplifies to counting the number of heading measurements aligned with forward direction $h_1$ and reverse direction $h_0$, and classifying the vehicle direction with the largest number of counts.

A description of the implementation follows. For vehicle objects initialized by a camera detection, the filter is unmodified: the object's relative heading is initialized to match the measurement $z_\theta$. However, for initialized objects, camera detection measurement updates of vehicles are modified; the measurement residual is computed for both the actual heading measurement $z_\theta$ and the reversed heading measurement $z_\theta + \pi$. The smaller residual measurement is applied and the respective count for heading aligned measurement $h_1$ or heading reversed measurement $h_0$ is incremented. After each heading measurement is applied to an object's KF, a classifier is run by simply selecting the maximum heading count classification $max(h_1,h_0)$. If the classifier finds the vehicle forward direction is reversed, then: the sign of the speed estimate is reversed, the heading angle is reversed 180 degrees, and the KF state covariance is updated accordingly.

\section{Experimental Hardware, Sensors, and Sensor Processing}
Experiments in this paper utilize Cornell's DUC entry vehicle, a 2007 Chevy Tahoe dubbed Skynet \cite{skynet}. The vehicle contains an assortment of radar, lidar, camera, GPS, odometry, and inertial measurement sensors. Internally, the vehicle features a 19U server rack for experimental algorithm deployment and data storage; experiment data can be collected for offline playback and evaluation. Typical operation with all sensors generates 350GB of raw uncompressed data per hour. Hardware upgrades since the DUC \cite{skynet} include a Point Grey Ladybug3 360 degree field-of-view spherical camera, external waterproofing of all sensor mounts and wiring, and upgrading of on-board computing and storage rack-mount servers.

\begin{figure} [h]
    \centering
    \includegraphics[width=6.5in]{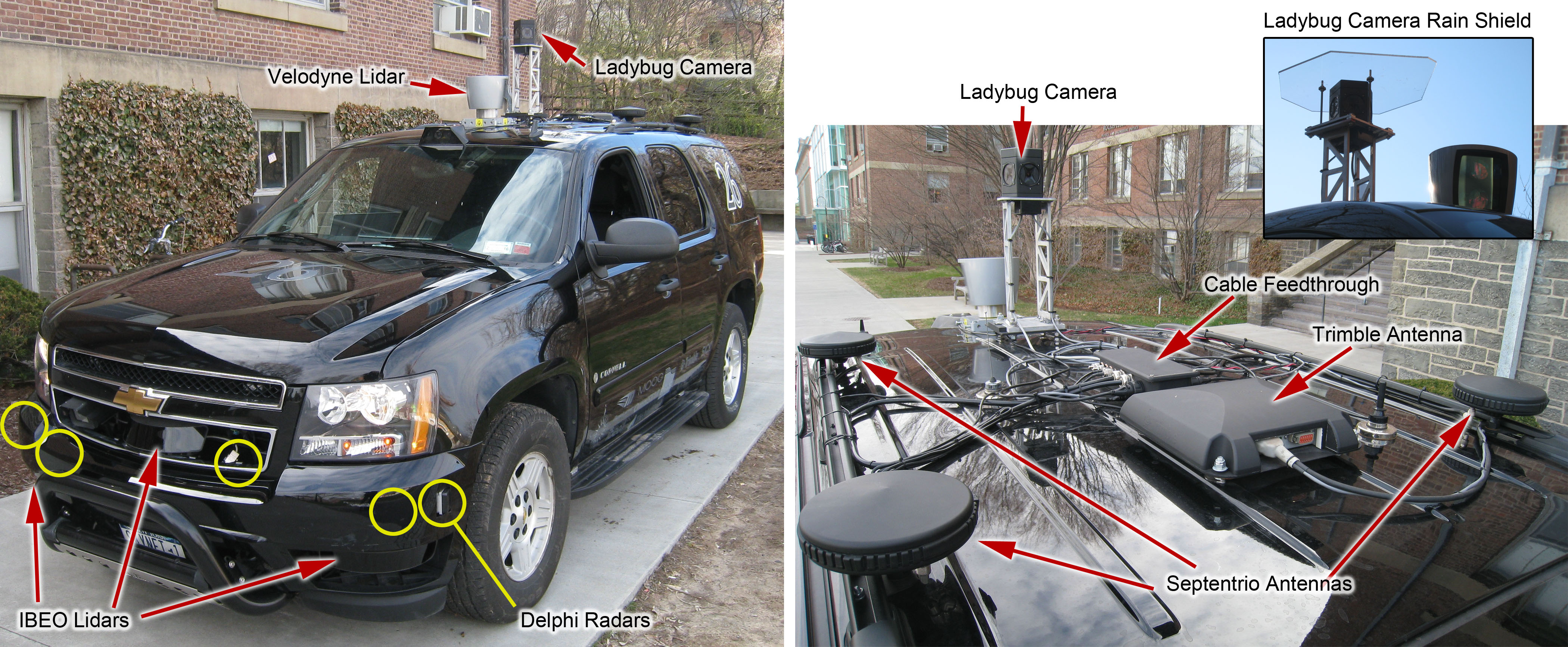}
    \caption{Skynet sensor suite.}
    \label{fig:skynet_exterior}
\end{figure}

Figure \ref{fig:skynet_exterior} shows the vehicle exterior with sensors well integrated with minimal external protrusions and wiring. Quality installation is critical to reliable performance in rain and snow. In addition to local environment perception sensors, Skynet is equipped with an attitude and position estimation system composed of a Litton LN-200 IMU, ABS wheel encoders, Septentrio PolaRx2e@ GPS receiver with three roof mounted antennas in an ``L'' configuration, and a roof mounted Trimble Ag252 GPS receiver. The LN-200 is floor mounted on the vehicle centerline above the rear axle. The Septentrio provides 5Hz synchronized GPS measurements of raw pseudo-range, Doppler shift, and carrier phase to satellites and decodes WAAS signal. The Trimble receiver decodes high-precision (HP) OmniSTAR differential corrections at 10Hz with 10cm accuracy. A pose estimator described in \cite{pose-skynet} combines sensor measurements to utilize strengths of each sensor and diversity to generate a robust attitude and position estimation solution.

\begin{figure} [h]
    \centering
    \includegraphics[width=3.2in]{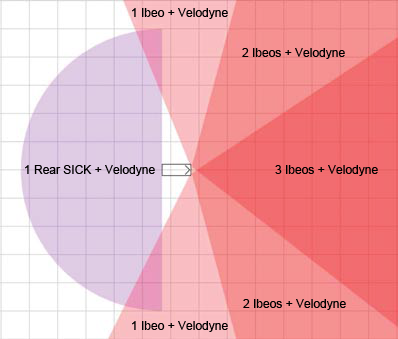}
    \includegraphics[width=3.2in]{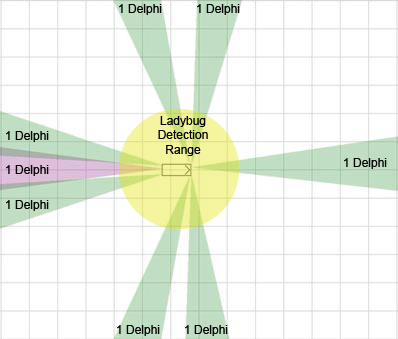}
    \caption{Sensor placement and coverage for lidar (left), camera (right), and radar (right) sensors. Skynet sits in the center of the diagram, facing right. Radar coverage is somewhat sparse in the forward direction. Original placement was chosen to detect objects along specific directions: forward for oncoming cars in opposing lanes, left and right for merging or intersection situations, and rear for passing situations. The lidar coverage is most dense in front of skynet. The Ladybug camera imagery detection range depends on resolution used for frame processing.}
    \label{fig:sensor_coverage}
\end{figure}

A 360 degree field-of-view Velodyne HDL-64E S1 lidar unit with 64 vertical laser scans is mounted on Skynet's roof. Velodyne lidar returns were used to detect the ground plane. Three IBEO XT Lidar units with 4 laser scans each are mounted on the front bumper. Ibeo lidar points are actively classified by Ibeo's proprietary software as object, rain, ground, or dirt, where dirt refers to lens cover fouling.

A Ladybug3 LD3-20S4C-33B spherical color camera is mounted behind and above the Velodyne for a clear view of surroundings. Narrow vertical struts provided rigid mounting and minimal blockage for the Velodyne lidar. The camera has five lenses pointed horizontally outward in a pentagon configuration and one lens pointed vertically. The camera lenses are factory calibrated to export 360 degree spherical or cylindrical projections with vendor provided software.  In lieu of building a complex mechanical wiper device, for experiments in this paper, a clear plexiglass cap shown in Figure \ref{fig:skynet_exterior} is mounted above the Ladybug camera to minimize lens fouling during precipitation.  Eight Delphi forward-looking millimeter-wave radar units are mounted around the vehicle; each includes proprietary black-box on-board processing to generate tracks in the form of object bearing, range, and radial speed. Tangential speed cannot be measured by radar and is not estimated by the radar unit's on-board processing. All sensor data is timestamped with 100usec accuracy via ethernet connected micro-controllers and a pulse clock synchronized to the LN-200 IMU. Figure \ref{fig:sensor_coverage} shows the sensor coverage; overlapping regions of Ibeo lidar and Ladybug camera in front of Skynet enable comparison studies between sensor modalities. The effect of processing resolution for Ladybug imagery is studied in the results section. The results sections use Ladybug imagery downsampled to one quarter resolution for near real-time detection processing due to computational constraints. Full resolution processing enables car detections up to 70m range and 40m range person detections; downsampling reduces the active range to 15m for cars and 10m for people. A 20m semicircle was used for quantitative evaluation comparisons.

Skynet is outfit with an occupancy grid as a safety catch-all to prevent running into objects that did not get tracked in the RBPF. This occupancy grid is a common commercial feature often implemented with sonar or radar. On Skynet this is implemented with lidar.

The classification formulations given in equations (\ref{eq:prob_class}) and (\ref{eq:heading_class}) can be numerically sensitive to machine precision underflow. Tracking errors induced from temporary filter instability, poor initial conditions, or a series of poor measurements can quickly drive a model's probability, for example $p(C=j|Z_k, A_k, O_k)$ or $p(H=h|Z_{k})$, to zero in machine-precision---that is, the incorrect model could have 100\% classification probability and the sequential estimator numerically multiplies future correct classification measurements by zero. The tracking filter may improve its estimate of object state, but the recursive classification multiplication is stuck at zero or one, effectively making an ad hoc classification due to machine precision underflow.  Implementation solutions include performing the probability calculation in log-likelihood units or thresholding the minimum or maximum classification probability values in the range $0+\epsilon<p (\cdot)<1-\epsilon$ for some small value of $\epsilon$, which was done in Skynet.

All perception algorithms aside from the camera detector run in real-time on a variety of Intel two and four core x86 64-bit processors in a Windows 7 environment. For simplicity, primary function algorithms are run on individual computers and data is shared between sensors and algorithms via UDP Multicast across a Hewlett-Packard  V1910-48G managed gigabit ethernet switch. The RBPF joint data association, tracking and classification routine is the most computationally intensive algorithm and reliably runs in real-time on an i7-3820 with 8 particles; that is eight complete hypothesis of the entire scene. Lidar clustering is run on an i7-930 while the remaining routines are run on an assortment of i5 and i7 processors. The following sections detail raw sensor processing, developed since the DUC \cite{skynet} and \cite{RBPF-Skynet}, to detect vehicles and people with the Ladybug camera and to process lidar returns for person-sized and car-sized clusters.

\subsection{Vision-based Detection}

The field of computer vision has been rapidly advancing. Recent studies have shown improved detection and classification rates for cars \cite{Held-ICRA2012} and pedestrians \cite{Angelova-2015}, \cite{Vasconcelos-Pedestrian}. The vision-based detection system presented below is a state-of-the-art detector for cars and pedestrians which also detects the heading of vehicles. Vehicle heading detection, a unique feature of the detector, is computed from each still frame from the camera, and thus does not require multiple frames or tracking through time. The following section explains how the Deformable Parts Model (DPM) technique \cite{Felzenszwalb-DPM} trained on existing datasets was used for both car and pedestrian detection.

The vision-based detection subsystem makes use of a Point Grey Ladybug 3 spherical camera mounted on top of the vehicle. Images are acquired at 6.5 frames per second synchronized with the vehicle's global clock and stitched into spherical panoramas using the vendor-provided software. Each spherical panorama is then reprojected into 8 separate rectilinear virtual cameras at 45 degree increments; reprojections are referred to as tiles.

Each tile is passed through two state-of-the-art detections, the first being a car detector and the second being a person detector. The car detector was first introduced in~\cite{nyc3dcars}. Both detectors make use of the Deformable Part Model (DPM), a technique for robustly detecting and localizing objects under varying viewpoint and illumination conditions by analyzing distributions of image gradients~\cite{Felzenszwalb-DPM}. Local gradient statistics are aggregated to form rigid \emph{parts}, small square patches that often have some semantic meaning, such as a wheel on a car. The model encodes the rest position of each part with respect to a root coordinate system, but additionally encodes an energy required to deform each part. For example, a single DPM can encode a large variety of car makes and models despite variations in shape or size by modeling how one part varies from a rest position according to the training set.

In addition to detecting and localizing passenger vehicles relative to the ego-vehicle, the car detector also predicts an orientation. A total of 16 separate DPM models are used to cover a set of orientations. Training examples are derived from three separate datasets: VOC2007~\cite{everingham-PASCAL10} -- an Internet dataset with 2D bounding box annotations, KITTI~\cite{Geiger2013IJRR} -- an autonomous vehicle dataset with 3D bounding box and orientation annotations fitted to lidar point clouds, and NYC3DCars~\cite{nyc3dcars} -- an Internet dataset with 3D bounding box and orientation annotations built by estimating scene geometry and asking annotators to place 3D models in the reconstructed scene. The detector from~\cite{voc-release5} is used for person detections.

After each tile has been processed, detections are aggregated per panorama and car orientation estimates are transformed from tile-space coordinates to vehicle-space coordinates. Non-maxima suppression is applied to threshold car and person detections. Detection boxes of pedestrians and cars are calculated in spherical angular coordinates. Object bearing relative to the ego vehicle is calculated from the centroid of the detection boxes. For autonomous vehicle tracking purposes, an estimate of object range is also helpful. By assuming the ego vehicle is oriented parallel to and contacting with the ground, and that all detected objects are in contact with the ground, a flat ground plane model and trigonometry can be used to estimate the nearest range in meters to the object and the object's width in meters.

The described algorithm is computationally intensive. An 8 core E5-2660 Xeon CPU and Nvidia GTX980 GPU were utilized to accelerate color processing, image resizing, and JPEG exporting of frames from recorded video sequences. Code implementation of the detection processing was performed on Amazon EC2 Cloud computing platform using 200 of their c3.large machine instances which contain 2 virtual CPU's (vCPU) with 3.75GiB of memory. The DPM is not optimized for computational efficiency or speed; thus, unoptimized implementation was run on a server farm. Full resolution panoramas from the Ladybug camera software were exported at 8000x4000 pixels, and tiles were exported at 2048x2048 resolution. Processing full resolution tiles through the DPM consumes all 4GB of RAM and takes 5.5 minutes per tile on a single machine. Reducing panoramas to 2000x1000 pixels and tiles to 512x512 drops computation time closer to real-time at 7 seconds per tile. A thorough analysis of imagery resolution and detection performance including recall and precision rates is included in the results section.

\subsection{Lidar Segmentation}
Car-sized clustering was unchanged from the DUC as described in \cite{skynet}. Raw lidar points, classified as objects by Ibeo's proprietary software, are further trimmed by removing any point within 0.3m of the detected ground plane. The remaining points are clustered into groups which have minimum size and maximum horizontal point spacing of 0.5m threshold and then a second time at a 1.0m threshold. Only objects that pass both clustering thresholds and whose maximum dimension is less than 15m (the maximum typical length of a bus) are considered `stable' clusters and passed to the joint perception algorithm. Clusters must contain at least seven points and must have at least one point projecting over 1.0m in height. Resulting clusters are classified as `car-sized'.

New person clustering is implemented by extending Laplacian of Gaussians (LoG) filtering typically used in image processing to lidar processing. Laplacian filters are derivative filters applied to images to find edges. Gaussian filters blur or smooth an image. For edge detection in traditional image processing, a normalized 2D LoG filter is built such that it calculates a large magnitude before and after an edge; there is a sign-change at the edge. With lidar, most objects appear as edges, but walking or standing people typically appear as a dense cluster of points separated from surrounding returns. In order to detect people, the filter is modified to find a person-sized group of points which are isolated from neighboring lidar returns. By constructing a normalized 1D LoG and sweeping the filter radially around a point, a 2d filter is constructed that has a negative mean value. The filter response is a negative value to a uniform field, a negative value to an edge, and a large positive value to a person-sized cluster isolated from its surroundings.

\begin{figure} [h]
    \centering
    \includegraphics[height=2.0in]{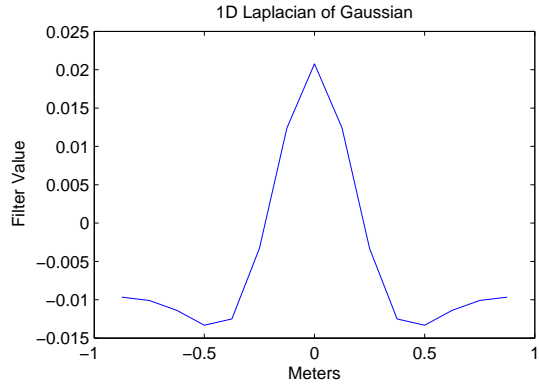}
    \caption{1D LoG filter for person lidar clustering.}
    \label{fig:LoG_kernel}
\end{figure}

\begin{figure} [h]
    \centering
    \includegraphics[height=2.0in]{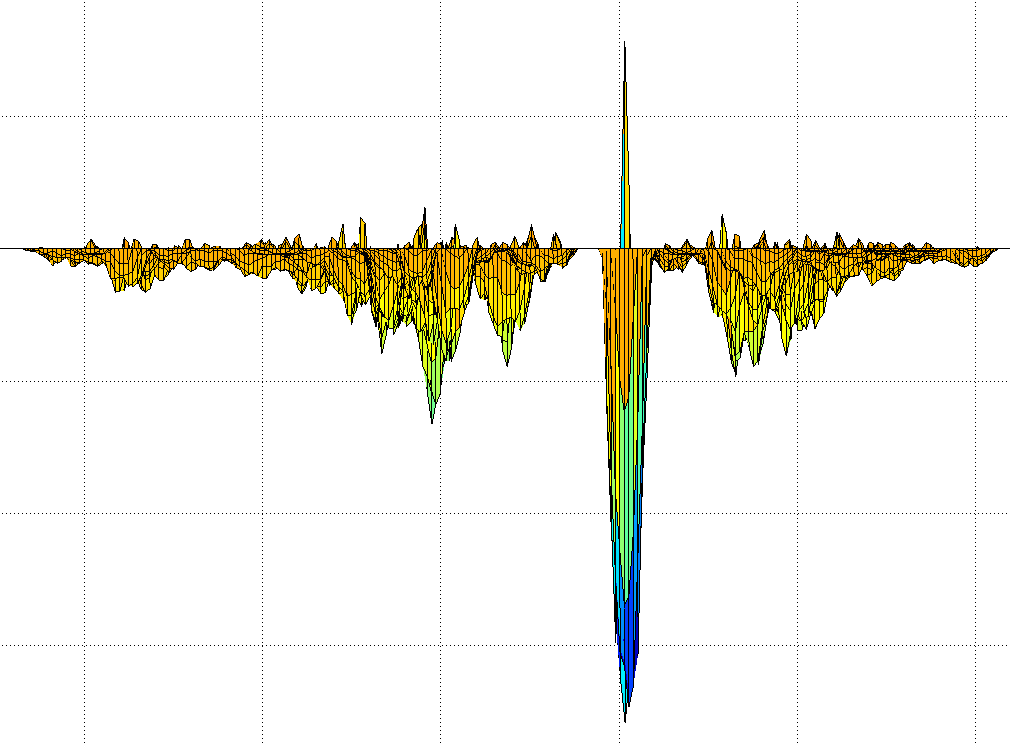}
    \caption{Example LoG filter response to lidar returns from an uncluttered roadway scene with one person.}
    \label{fig:LoG_returns}
\end{figure}

Object point returns are projected into a horizontal plane and grouped in a square grid with 25cm sized cells. The heuristically modified LoG response is computed per lidar return 
\begin{equation}
\begin{aligned}
L(r)=\frac{1}{\pi\sigma^2}\left[1-\frac{r^2}{2\sigma^2}\right]e^{\frac{-r^2}{2\sigma^2}} + \frac{0.15}{\pi\sigma^2}\left[1-\frac{r^4}{2\sigma^2}\right]e^{\frac{-r^2}{2\sigma^2}}
\end{aligned}
\end{equation}
where $r$ is the radius from filter center in the horizontal plane and $\sigma=0.45$m. A plot of the filter magnitude versus radius is shown in Figure \ref{fig:LoG_kernel}. By computing the filter response at each grid cell as $\sum L(r)$ for all lidar returns, a person cluster can be identified. Figure \ref{fig:LoG_returns} shows an example uncluttered roadway scene containing 1 person after being processed by the LoG filter for a person. The positive peak is the person's location; the negative peak is the area neighboring the person; and other negative areas correspond to other objects and edges from the scene. Non-maximum suppression and a minimum positive signed amplitude threshold is applied across the grid to select peak locations for person-sized clusters. Both person-sized and car-sized lidar clusters are then passed to the joint association, tracking, classification algorithm. Recall and precision rates are included in the results section.

\section{Simulation and Experimental Results}

The results section of this paper first motivates probabilistically adding classification to the tracking problem. Kalman Filter inference of object classification is demonstrated in both Monte Carlo simulation and experimental data by running a bank of filters on object measured data. The experimental portion uses data collected with a low accuracy hand-held GPS unit from pedestrian, biker, car, and bus objects traversing around Cornell's campus.

Next, the joint data association, tracking, and classification algorithm performance is evaluated using Skynet---Cornell's AGV entry from the DUC. Repeatable scenarios of intersection encounters were conducted in multiple weather conditions and recorded with camera `C', lidar `L', and radar `R' sensors. Quantitative evaluations were conducted in post-analysis with the reduced sensor sets L+R and C+R and complete sensor set C+L+R in order to evaluate if cameras can replace lidars for AGV applications. Performance of vision heading detection inclusion in the estimator and selection of the number of particles used in the RBPF are analyzed. Ground truth of two pedestrians and one vehicle enables quantitative evaluation of tracking and classification performance. The joint solution demonstrates robustness to all weather conditions.

Performance of reduced sensor set evaluations from the quantitative experiments provides insight into individual sensor performance in different weather conditions in staged scenarios. Additional non-staged qualitative experiments were conducted by driving Skynet through traffic around downtown Ithaca, NY to more broadly evaluate sensor and perception algorithm performance in varied weather conditions.

\subsection{Monte Carlo Simulations: Joint Tracking-Classification}

Monte Carlo simulations were conducted to evaluate classification performance of two categories, person and cyclist, given truth data generated with a dynamics function that exactly matches the dynamics modeled in the KF.

\begin{figure} [h]
    \centering
    \includegraphics[height=3in]{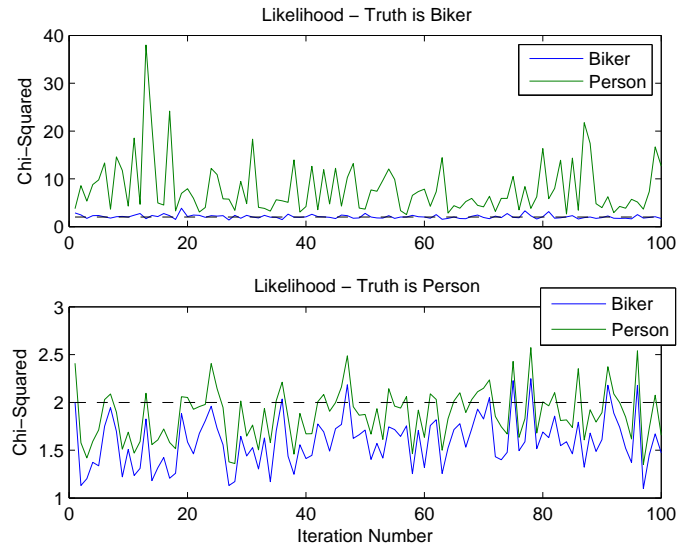}
    \caption{Monte Carlo simulation results. Black dashed line represents ideal mean value for true classification.}
    \label{fig:MC-sim}
\end{figure}

Synthetic tracks, 50 seconds in length, sampled at 1Hz, of both people and cyclists were generated from the KF modeled dynamics functions. Motion occurred from randomly selected initial conditions and process noise; synthetic measurements were created by adding Gaussian measurement noise to the synthetic tracks. Measurement noise was randomly drawn per track inside the MC truth simulation; each track is considered an MC iteration. Measurements from each of these tracks were then run through both KFs and classified based on the highest $\chi^2$ probability. Both filters are initialized by setting the initial position equal to the first position measurement and the initial velocity vector tangent to the line connecting the first and second position measurement. The limit of the $50^{th}$ percentile $\chi^2$ cumulative density function (cdf) should approach the number of degrees of freedom in the measurement vector which is two. By averaging the $\chi^2$ values over multiple KF updates, a distribution with $n_{\tt{DOF}}$ degrees of freedom is generated as follows
\begin{equation}
\begin{aligned}
n_{\tt{DOF}} = kn_z
\end{aligned}
\end{equation}
 where $k$ is the number of measurements and $n_z$ is length of the measurement. For the MC simulations, $k=50$ measurements of dimension $n_z=2$ provides a distribution with $n_{\tt{DOF}}=100$ degrees of freedom. The  $50^{th}$  percentile $\chi^2$ cdf can be calculated as follows
\begin{equation}
\begin{aligned}
\chi^2_{\tt{cdf}} &= \chi^2_{\tt{cdf}}({\tt{percentile}},n_{\tt{DOF}})/n_z\\
&= \chi^2_{\tt{cdf}}(0.5, 100)/50\\
&= 99.3/50=1.986
\end{aligned}
\end{equation} 
This $50^{th}$ percentile is shown as a black dashed line in Figure \ref{fig:MC-sim} along with the $\chi^2$ averages for each test run. The green trace is the likelihood the track was generated by a person while the blue trace is the likelihood the track was generated by a biker. The top plot track iterations were generated with cyclist wheeled dynamics while the lower plot track iterations were generated with the person walking dynamics. 

Quantitative numbers summarizing the plot are show in Table~\ref{kfclass-mc-results}. The $\chi^2$ average listed in the table is the average value of the traces from Figure~\ref{fig:MC-sim} over all 100 iterations. The true model classification averages close to the ideal 2.0 mean while the incorrect model is farther from 2.0; these results show that, given enough data and the models, the biker and person can be inferred. The simulation correctly classified 100\% of cyclist and 79\% of the person tracks.

\begin{table}[h]
\caption{Monte Carlo simulation results.} \label{kfclass-mc-results}
\begin{center}
\begin{tabular}{|l|c|c|c|c|}
  \toprule
 \multirow{2}{*}{Track Generation:} & \multicolumn{2}{c|}{Average $\chi^2$} & \multicolumn{2}{c|}{$\chi^2$ Classification Percentage} \\
 & Cyclist & Pedestrian & Cyclist & Pedestrian \\
  \midrule\midrule
  Cyclist & 2.0897 & 7.9631 & 100 & 0 \\
  Pedestrian & 1.5916 & 1.8784 & 21 & 79 \\
  \bottomrule
\end{tabular}
\end{center}
\end{table}

\subsection{Experimental Results: Joint Tracking-Classification}

Experiments studying joint tracking and classification were conducted by building KF models of four classifications: pedestrian, biker, car, and bus. A low accuracy hand-held GPS with 1.1 meter standard deviation of error recorded position data from each object. A portion of the data was used to estimate the process noise parameters for each class. Classification performance was analyzed on the remaining data to demonstrate that accurate classification can be accomplished using only GPS position measurements of the objects.

All data was collected in Ithaca, NY around Cornell University campus and the Ithaca Commons. Euler integration is used to predict the model of the continuous time obstacle dynamics at 1Hz intervals. GPS data was collected using a Locosys GT-31 hand-held unit at 1Hz frequency and recorded in NMEA-GGA sentence format, which has approximately 18cm of quantization error in Ithaca, NY. Measurement noise for the sensor was computed using data collected from a stationary sensor to establish a covariance matrix $R$ in meters squared as follows
\begin{equation}
\begin{aligned}
R=\begin{bmatrix}1.2 & 0.1 \\ 0.1 & 1.2\end{bmatrix}.
\end{aligned}
\end{equation}
Car data was collected using a 4-door sedan while bus data was collected while riding on a local commuter bus. The filters are not expected to distinguish non-moving targets. Because some of the driving data contains time waiting at traffic lights, any data sections where the GPS unit moved less than 25cm on average between downlinks was removed from the evaluation set. A holdout set of the data 5 minutes in length was used with a Matlab minimization routine to estimate the process noise covariance values for the four models and is given in Table~\ref{dyn-proc-noise}.

\begin{table}[h]
\caption{Process noise values.} \label{dyn-proc-noise}
\begin{center}
\begin{tabular}{|r|c|c|c|c|}
  \toprule
   Model: & Pedestrian & Car & Bus & Cyclist \\
  \midrule\midrule
  Acceleration $\tt{m/s^2}$ & 0.04 & 0.6 & 0.4 & 0.31 \\
  Rotation Rate $\tt{deg/s}$ & N/A & 15 & 15 & 15 \\
  \bottomrule
\end{tabular}
\end{center}
\end{table}

\begin{table}[h]
\caption{Successful classification rate from experimental GPS data.} \label{kfclass-exp}
\begin{center}
\begin{tabular}{|r|c|c|c|c|}
  \toprule
   Model: & Pedestrian & Car & Bus & Cyclist \\
  \midrule\midrule
  Percentage & 100\% & 99\% & 99\% & 50\% \\
  \bottomrule
\end{tabular}
\end{center}
\end{table}

From each experimentally recorded data set, 100 sections of 50 consecutive GPS data points were randomly selected and fed into the bank of Kalman filters enumerating each classification. Classification was inferred using all four KFs and reported in Table~\ref{kfclass-exp}.  Pedestrian, car, and bus were correctly classified for at least 99\% of the tested tracks by the $\chi^2$ statistics. The cyclist classification rate was much lower, in part because the collected cyclist data contained more stops and starts; of the three other classes, the cyclist was most often mis-classified as a pedestrian. In summary, the example with experimental data shows reliable classification performance solely from position data, highlighting an example benefit possible with joint tracking and classification solutions.

\subsection{Experimental Data Collection}

Several datasets were collected in order to study key elements of the solution, including algorithm parameters, sensor types, and weather. In the first dataset, experiments were conducted with specified encounters on a closed course of one vehicle and two pedestrians and were repeated multiple times in different weather and lighting conditions to understand how weather affected sensor and algorithm performance.

Quantitative analysis of the controlled experiments is enabled by using a 1999 Chevrolet Suburban as a ground truth vehicle. The Suburban is outfit with the same GPS and inertial measurement sensors as Skynet, which resolve centimeter-accuracy position estimation reported at 100Hz of the vehicle obstacle. By differencing high precision pose estimates of the Suburban from Skynet, a relative truth dataset was obtained for the various encounters. Real-time sub-meter GPS sensing for pedestrians was not available. Instead a combination of a pre-surveyed path, low precision GPS, and a camera tracker were used. The pre-surveyed path was marked for pedestrians in the closed course. The time-synchronized camera recorded accurate timing of the pedestrian's location on the surveyed path, enabling generation of sub-meter accuracy truth data for the pedestrians in the closed course. All truth data is synchronized in post-processing using GPS recorded timing information.

\begin{figure} [h!]
    \centering
\begin{tabular}{ccc}
    \includegraphics[height=2.5in]{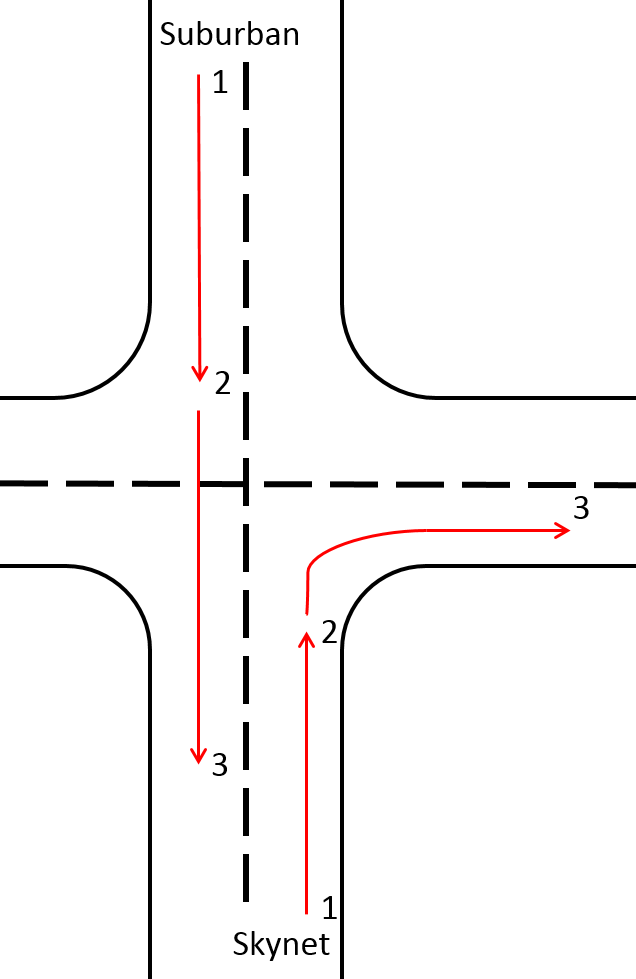}&
    \includegraphics[height=2.5in]{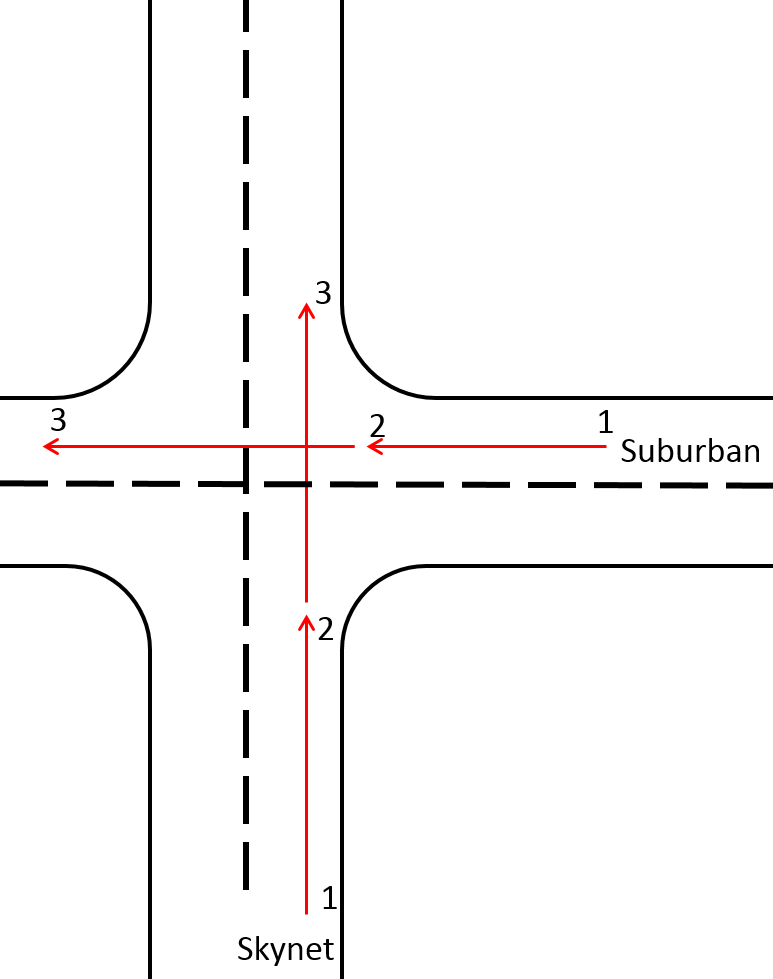}&
    \includegraphics[height=2.5in]{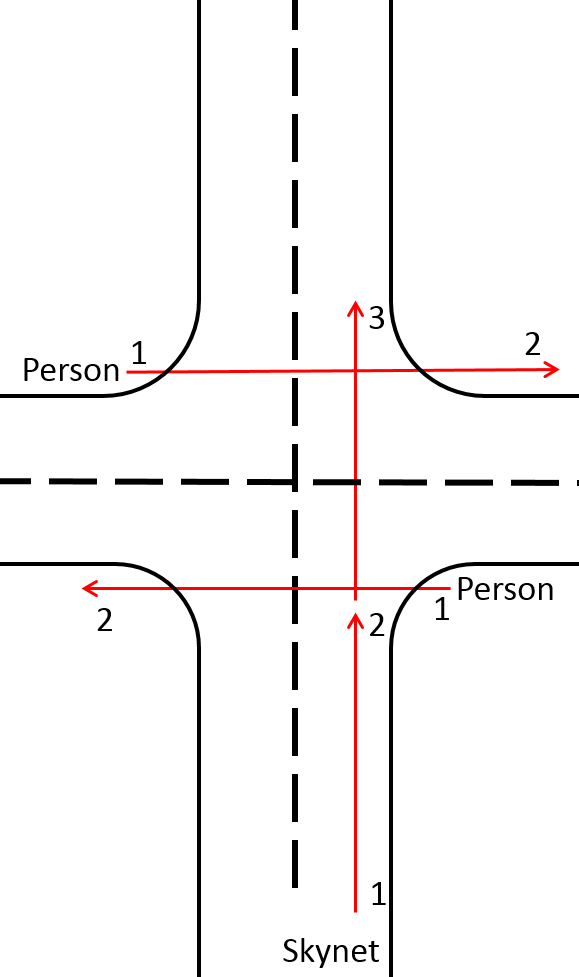}\\
    Scenario A&Scenario B&Scenario C
\end{tabular}
    \caption{Three intersection scenarios used in the experiments are defined as shown.}
    \label{fig:exp-trials}
\end{figure}

\begin{figure} [h]
    \centering
    \includegraphics[height=4.0in]{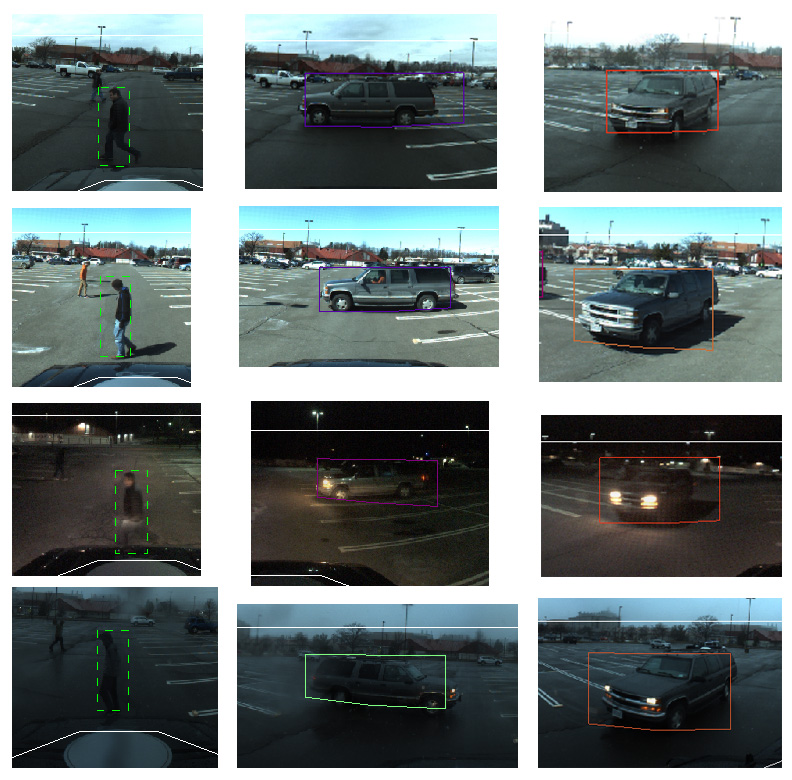}
    \caption{Visual data and example detections for the four weather condition categories during the intersection experiments. Rows top to bottom: Cloudy \& Wet, Sunny, Night, Snow \& Rain.}
    \label{fig:weather-conditions}
\end{figure}

Three experimental scenarios in Figure \ref{fig:exp-trials} quantitatively evaluate the repeatability and performance of the RBPF across four different weather scenarios. For Scenario A, the Suburban and Skynet drive towards each other and stop at an intersection; Skynet turns as the Suburban drives straight through the intersection. For Scenario B, Skynet and the Suburban approach and stop at an intersection, and the Suburban crosses in front of Skynet. For Scenario C, Skynet approaches and stops at an intersection, and two pedestrians cross in opposite directions in front of Skynet; Skynet proceeds through the intersection after pedestrians cross. For each scenario five experimental trials were conducted for each weather condition. The experiment was conducted over four different weather condition categories: Sunny, Night, Wet \& Cloudy, and Snow \& Rain. Visual data and example detections for all four weather condition categories are shown in Figure~\ref{fig:weather-conditions}. The Wet \& Cloudy trials were recorded after a rain storm; the ground was wet but no precipitation was present and the sun was occluded by clouds. This condition most closely resembled the conditions on the second day of the KAIST competition that resulted in two autonomous vehicles crashing \cite{KAIST-weather}.

\begin{figure} [h]
    \centering
    \includegraphics[width=4.0in]{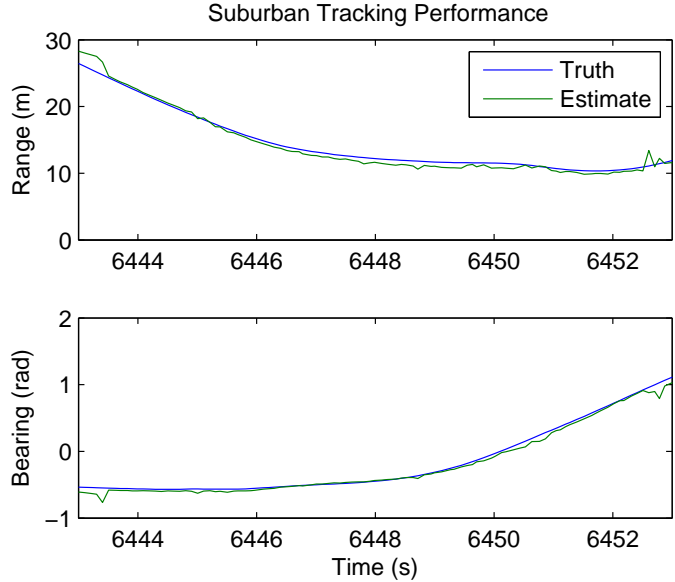}
    \caption{Example truth and estimated object track for Suburban.}
    \label{fig:suburban-tracking}
\end{figure}

By logging all timestamped sensor recordings, data could be replayed through the algorithms for studies such as a reduced set of sensors or modifications to the RBPF. Quantitative evaluation is achieved by using the most likely particle from the RBPF and comparing the closest tracked object to the truth object, which could be a pedestrian or Suburban depending on the scenario. The closest object to truth is considered correctly tracked if the truth's and estimate's object centroids are within 2m. The RBPF output is recorded at 10Hz for evaluation in a 20m semi-circle in front of Skynet. The results tables in this section contain columns with the following specific definitions: Object Tracked is the fraction of time an estimated object from the RBPF overlapped the truth location; Range and Bearing Errors are reported as root-mean-square (rms) statistics of the range to closest point and mean bearing respectively; Correct Classification is the fraction of time that a tracked object was correctly classified, while Mis-Classification is the fraction of time a pedestrian was errantly classified as a car or vice versa; and Number of Returns is the number of time instances that the estimated object overlapped the truth object. Unclassified time fraction is not reported in the tables but may be easily calculated; the sum of Correct, Mis-, and Un- classified must equal one. An example encounter of an object being properly tracked in range and bearing is shown in Figure \ref{fig:suburban-tracking}.

In addition to the controlled dataset, an additional experiment in an uncontrolled environment in Sunny conditions was conducted by driving Skynet through Cornell's campus from B-Lot past the Engineering Quad through CollegeTown for 7 min 13.4 sec. This experiment contained a high density of pedestrian and vehicle traffic and is used in the following results sections to compare vision resolution performance, sensor precision and recall rates, and particle count selection for the RBPF filter. Truth data was unavailable for this uncontrolled experiment.

\subsection{Particle Count Selection for Joint Data Association, Tracking, and Classification}

\begin{table}[htbp]
    \centering
  \caption{RBPF performance for different numbers of particles. \emph{For all table results:}  Object Tracked is the fraction of time an estimated object from the RBPF overlapped the truth location; Range and Bearing Errors are reported as root-mean-square (rms) statistics of the range to closest point and mean bearing respectively; Correct Classification is the fraction of time that a tracked object was correctly classified, while Mis-Classification is the fraction of time a pedestrian was errantly classified as a car or vice versa; and Number of Returns is the number of time instances that the estimated object overlapped the truth object. Unclassified time fraction is not reported in the tables but may be easily calculated; the sum of Correct, Mis-, and Un- classified must equal one.} 
     \resizebox{\linewidth}{!}{\begin{tabular}{| >{\centering\arraybackslash}p{2cm} | >{\centering\arraybackslash}p{1.25cm} | >{\centering\arraybackslash}p{1.8cm} >{\centering\arraybackslash}p{2.0cm} | >{\centering\arraybackslash}p{2.2cm} >{\centering\arraybackslash}p{2.2cm} | >{\centering\arraybackslash}p{2cm} |}
    \toprule
    {\textbf{Number of Particles}} & {\textbf{Object Tracked}} & {\textbf{Range \mbox{Error} (m)}} & {\textbf{Bearing Error (rad)}} & {\textbf{Correct \mbox{Classification}}} & {\textbf{Mis-Classification}} & {\textbf{Number of Returns}} \\
    \midrule\midrule
    {\textbf{1}} & {0.993} & {1.698} & {0.068} & {0.917} & {0.001} & {1484} \\
    {\textbf{4}} & {0.998} & {1.545} & {0.076} & {0.870} & {0.000} & {1496} \\
    {\textbf{8}} & {0.998} & {1.544} & {0.077} & {0.875} & {0.000} & {1455} \\
    \bottomrule
    \end{tabular}}%
  \label{tab:results_particles}%
\end{table}%

A key design consideration for the RBPF is selecting an appropriate number of particles. Each particle represents a full hypothesis of the measurement associations and object states in the local environment. In Miller's original RBPF \cite{RBPF-Skynet}, multiple hypotheses helped model ambiguity associated with data association. For Cornell's DUC entry, a total of four particles ran in real-time and adequately captured variability in data association, primarily due to the wide spacing between vehicles. With the addition of classification, the RBPF can help model ambiguity regarding object type classification, such as car or pedestrian, or capture other binary object characteristics such as a vehicle's true heading direction.

For more complex scenes, such as those considered here with both cars and pedestrians, the number of particles may need to be higher. The first study of particle counts uses the sunny dataset; all trials from all scenarios were used, and the RBPF was run with 1, 4, and 8 particles to study the effect on performance. Results in Table~\ref{tab:results_particles} show negligible performance improvement comparing 1, 4, and 8 particles. The association ambiguity was negligible for the largely spaced objects in these intersection scenarios; hence performance was similar across the different number of particles, and controlled intersection experiments could be analyzed with only a single particle.

The second particle study used the data recorded driving through Cornell's campus and CollegeTown, a more complex scene with numerous pedestrians and vehicles making data association and classification less obvious. Truth data was unavailable for this uncontrolled experiment. Data is played back with two configurations of algorithms: 1) Miller's original Clustering and RBPF algorithms from the DUC, and 2) the extended Clustering and RBPF with classification algorithms presented in this paper. Comparisons between the two configurations help understand how selection of RBPF particle count is affected by the addition of classification. Given the RBPF is extended to classify two object types, namely cars or pedestrians, it was initially hypothesized that doubling the number of particles used in the original DUC RBPF from four to eight might provide adequate performance. Data is passed to both algorithm configurations and run with 1, 4, 8, 12, 16, and 20 particles. A playback using 50 particles for each configuration is treated as the benchmark for comparison. Errors in object counts are computed by differencing individual runs against the respective configuration's 50 particle count run.

\begin{figure} [h]
    \centering
    \includegraphics[width=\linewidth,trim={2cm 0 2cm 0},clip]{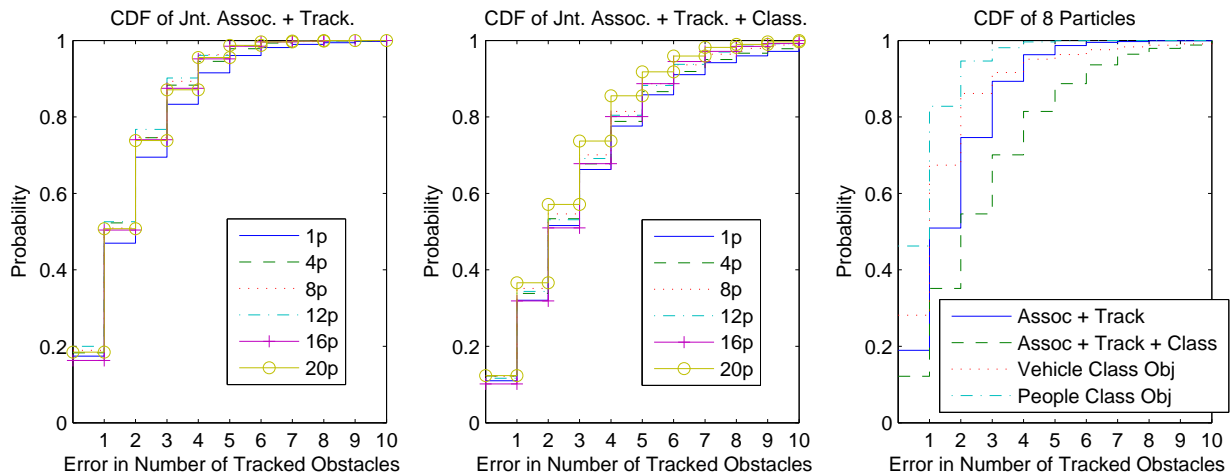}
    \caption{Comparison of number of particles selected for joint data association and tracking versus joint data association, tracking, and classification. Figures plot cumulative distribution function (CDF) of errors in the number of tracked objects for various particle counts versus a truth simulation using 50 particles. The right-most plot compares how the addition of classification affects the overall object error CDF for 8 particles. The added complexity, increased number of measurements, and addition of classification decreases the overall CDF, implying more particles are necessary for similar performance when adding classification. However, CDF errors for car and person classified objects is decreased, that is, classification improved the object track consistency for classified objects compared to omitting classification.}
    \label{fig:CDF-particles}
\end{figure}

\begin{table}[h]
\centering
\caption{Joint Data Association and Tracking (DUC Tracker)} \label{tab:Particle-stats}
\begin{tabular}{|>{\centering\arraybackslash}m{2cm} | >{\centering\arraybackslash}m{3.5cm}| >{\centering\arraybackslash}m{3.5cm}| >{\centering\arraybackslash}m{3.5cm}|}
  \toprule
 & \textbf{Joint Association \& Tracking} & \multicolumn{2}{c|}{\textbf{Joint Association, Tracking, \& Classification}}\\
\midrule
   \textbf{Number of Particles} & {\textbf{RMS Error: Number of Objects}} & \textbf{RMS Error: Number of Objects} & \textbf{RMS Error: Num. Classified Objects} \\
  \midrule\midrule
  1 & 2.76 & 4.39 & 1.94 \\
  4 & 2.45 & 4.10 & 1.80 \\
  8 & 2.32 & 3.80 & 1.80 \\
  12 & 2.21 & 3.57 & 1.84 \\
  16 & 2.33 & 3.54 & 1.64 \\
  20 & 2.31 & 3.18 & 1.78 \\
  \bottomrule
\end{tabular}%
\end{table}%

\begin{figure} [h]
    \centering
    \includegraphics[width=3.0in]{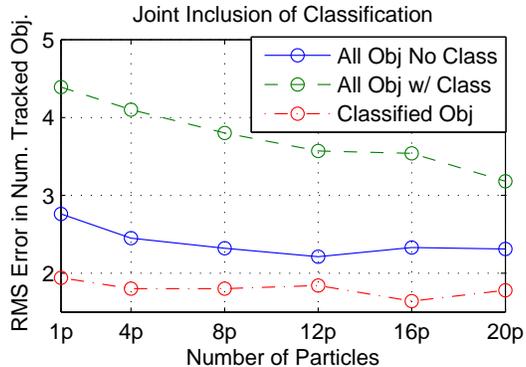}
    \caption{Comparison of overall errors in the number of tracked objects. This graphic plots data from Table~\ref{tab:Particle-stats}.}
    \label{fig:Particle-stats}
\end{figure}

Table~\ref{tab:Particle-stats} and Figure~\ref{fig:Particle-stats} show the results from Miller's DUC tracker (configuration 1) which used only car-sized lidar clusters, no Ladybug imagery, and no classification; and the joint classification-tracker developed here (configuration 2) including clustering, imagery, and joint data association, tracking, and classification routines. The truth simulation run tracked an average number of 34 objects for configuration 1 and 66 objects for configuration 2, of which on average 8 were classified as cars and 4 were classified as pedestrians. Plots of the cumulative distribution function (CDF) of errors in the number of tracked objects is presented in Figure~\ref{fig:CDF-particles}.

For configuration 1, joint association and tracking, in accordance with \cite{RBPF-Skynet}, increasing particle count from 1 to 4 improves performance; performance plateaus at 8 particles. For configuration 2, adding classification along with vehicle heading ambiguity, pedestrian lidar clusters, and increased number of camera returns, provides significant additional complexity. The RBPF filter continues to improve performance with increased number of particles out to 20 particles, as highlighted in Figure~\ref{fig:Particle-stats}. Put another way, errors in overall (classified + non-classified) object counts decrease with more particles. However, errors in the number of classified objects plateaus at 4 particles, so the addition of classification required a minimum of 4 particles. Objects that have been classified are well-tracked; higher particle counts provided negligible benefit for these well-tracked objects. Four particles is adequate for classified objects because it is rare for an object to simultaneously have ambiguous data association, classification, and heading direction. The benefit in increasing to a higher number of particles was observed for objects that do not get classified such as shrubbery, buildings, or distant pedestrians/vehicles that have sparse measurement returns.

Configuration 2 tracks a larger overall number of objects than configuration 1 which contributes to configuration 2's higher overall error rate of number of tracked objects. However, the error in number of classified objects tracked in configuration 2 is less than the error in number of unclassified objects in either configuration; thus, classification helps to improve the overall perception system's data association and tracking performance.

The number of reasonable and therefore possible associations and object classifications can increase due to closely spaced objects like pedestrians, which makes selection of an optimal number of particles non-obvious. When considering more than 4 particles, a trade-off is reached between increased computational complexity and probabilistic fidelity. In general, the results of classified object track count errors plateauing at 4 particles implies that 4 particles is adequate for classified object tracking in urban driving scenarios, but additional particles support more congested scenes. For the other studies in this paper, 8 particles was chosen due to its balance between performance and real-time capability in the author's C++ implementation. All remaining experiments, both controlled intersection scenarios and urban downtown driving, are run with 8 particles.

\subsection{Controlled Experiments: Sensor Sets in Joint Data Association, Tracking, and Classification}

\begin{figure} [h]
    \centering
    \includegraphics[width=6.5in]{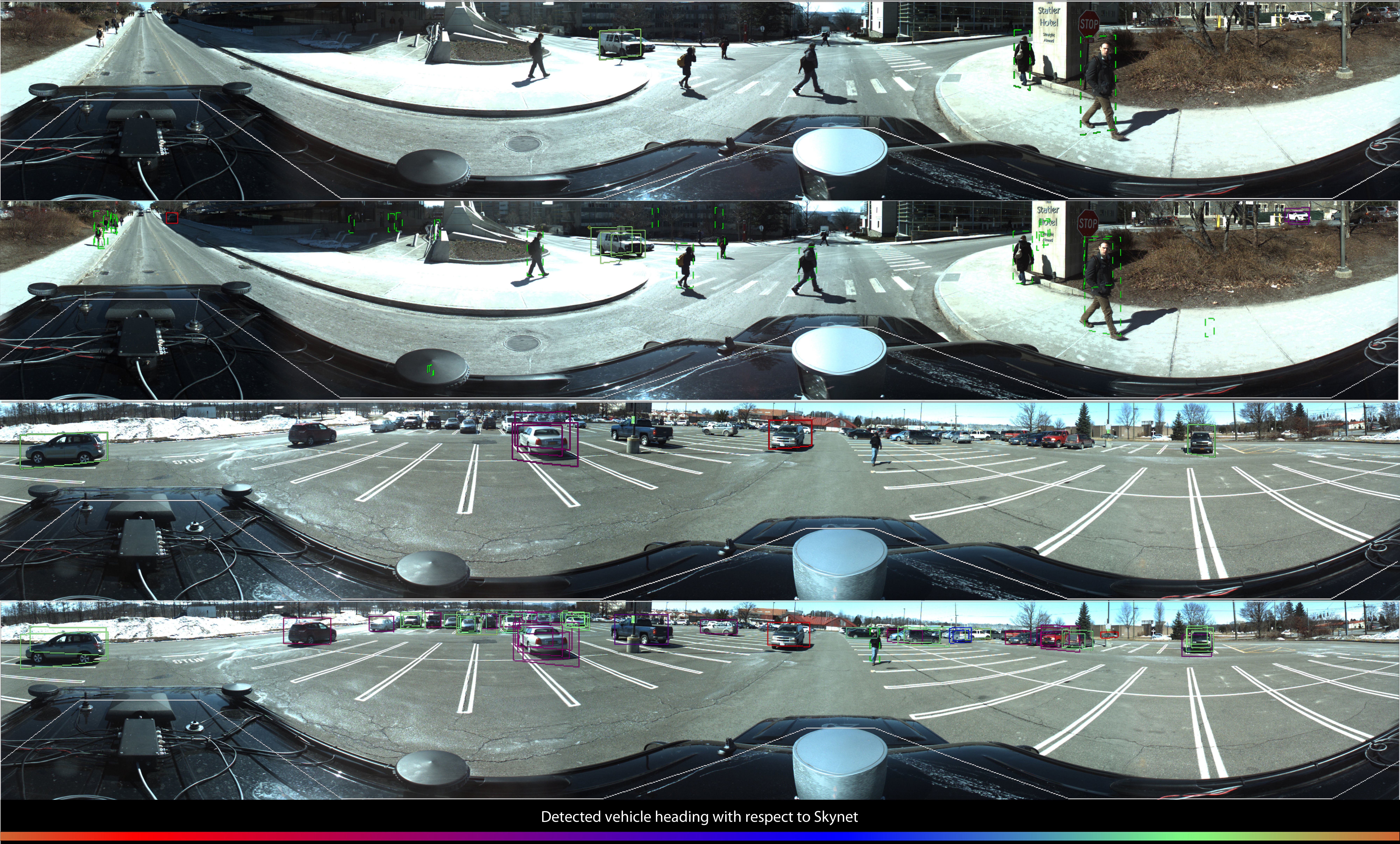}
    \caption{Example of detection performance of the car and person detector for the same images but different processing resolutions. The second and fourth photos show detections for 8k resolution imagery from Ladybug camera. The first and third photos show car and person detections for 2k imagery used for remaining experiments.}
    \label{fig:detection_example}
\end{figure}

\begin{table}[h]
\centering
\caption{Camera processing per resolution. For reference, 2k is similar in resolution to 1080p HDTV.} \label{tab:Cam-Resolution}
\begin{tabular}{|l|c|c|c|c|}
  \toprule
    & \textbf{1k} & \textbf{2k} & \textbf{4k} & \textbf{8k} \\
  \midrule
Actual Resolution: & 1024x512 & 2048x1024 & 4096x2048 & 8000x4000 \\
Tile Resolution: & 256x256 & 512x512 & 1024x1024 & 2048x2048 \\
Car Detection Range: & 7m & 15m & 32m & 70m \\
Person Detection Range: & 5m & 10m & 20m & 40m \\
  \bottomrule
\end{tabular}%
\end{table}%

\begin{table}[h]
\centering
\caption{Sensor recall and precision rates. Recall is computed over sensor's range.} \label{tab:Recall-Precision}
\begin{tabular}{|l|c|c|c|c|}
  \toprule
   \textbf{Sensor, Condition} & \textbf{Car Recall} & \textbf{Car Precision} & \textbf{Person Recall} & \textbf{Person Precision} \\
  \midrule\midrule
  Lidar, Sunny & 91\% & 51\% & 88\% & 55\% \\
  Camera, Sunny & 85\% & 95\% & 70\% & 90\% \\
  Camera, Rainy & & 93\% & & 76\% \\
  \bottomrule
\end{tabular}%
\end{table}%

\begin{table}[ht]
\centering
\caption{Sensor detection rates averaged per sensor return.} \label{tab:Detection-Rate}
\begin{tabular}{|l|c|c|c|c|}
  \toprule
   \textbf{Sensors Used} & \textbf{Objects Detected} & \textbf{Cars Detected} & \textbf{Persons Detected} \\
  \midrule\midrule
C-1k & 0.61 &  0.46 & 0.15 \\
C-2k & 1.98 & 1.53 & 0.45 \\
C-4k & 10.50 & 4.52 & 5.98 \\
C-8k & 35.44 & 8.12 & 27.31 \\
L & 12.04 & 9.26 & 2.78 \\
  \bottomrule
\end{tabular}%

\centering
\caption{Sensor object tracking and classification average rates.} \label{tab:Tracking-Rate}
\begin{tabular}{|l|c|c|c|}
  \toprule
   \textbf{Sensors Used} & \textbf{Objects Tracked} & \textbf{Cars Tracked} & \textbf{Persons Tracked} \\
  \midrule\midrule
C-1k + R & 41.10 & 3.00 & 0.04 \\
C-2k + R & 42.86 & 7.31 & 0.22 \\
C-4k + R & 50.02 & 14.14 & 2.49 \\
C-8k + R & 63.16 & 27.99 & 4.35 \\
L + R & 63.11 & 0.27 & 3.20 \\
C-8k + L + R & 81.86 & 27.67 & 9.44 \\
  \bottomrule
\end{tabular}%
\end{table}%

Examination of sensor segmentation performance, along with the output's integration into the joint association, tracking, and classification algorithm is presented in this section. The first study is the performance of the vision detection algorithm as a function of image resolution. As described in the computer vision section, camera images were downsampled due to computational restrictions. Table~\ref{tab:Cam-Resolution} presents four levels of resolution, along with the observed car and pedestrian detection ranges. Full resolution 8k video is reduced to the near real-time version of 2k which contracted the car detection range from 70m to 15m and person detection range from 40m to 10m. A comparison of full and reduced resolution detections from the Ladybug camera is shown in Figure~\ref{fig:detection_example} for two scenes. In the upper photo pair, 2 people are detected in the downsampled image versus 13 in the full resolution image. In the lower photo pair, 4 cars are detected in the downsampled image versus 26 in the full resolution image. Downsampling images reduces the number of pixels describing an object; given enough downsampling any object will become obscured. In the shown images, smaller and more distant objects are not detected in the downsampled versions. In practice, reducing camera resolution reduces object detection range.

Using a holdout set of data, recorded from driving 3.0 miles from Cornell's campus through College Town to downtown Ithaca, NY, true classification rates of precision were computed for car and person detection routines by hand labeling at least 300 detections. Summary statistics  of recall and precision are presented in Table~\ref{tab:Recall-Precision}. When normalized by sensor range, recall and precision rates were found to be independent of camera resolution. During clear daytime conditions, camera detections had a correct classification rate of 95\% for a car and 90\% for a person. During rainy conditions, camera true classification rates rates dropped to 93\% for car and 76\% for pedestrian. Thus, one car detection or two person detections are sufficient for a 95\% confident classification of a respective car or person object. Clearly, as computation becomes faster over time, the performance of the visual detector---particularly in range---will improve dramatically. Recent developments such as \cite{DPM-30Hz}, \cite{Angelova-2015}, and \cite{Vasconcelos-Pedestrian} have demonstrated real-time implementation by intelligently selecting regions of interest for processing and by utilizing GPU acceleration for DPM implementation.

True detection rates (precision) for person and car classified lidar clusters are calculated by analyzing performance of clustering across clear day and night weather scenarios using a holdout dataset collected from Cornell campus, College Town, and downtown Ithaca, NY. Lidar clustering maximum range is approximately 70m. The lidar clustering routines are designed based on object size, not object feature extraction; thus similarly sized non-person and non-car objects are clustered as person-sized or car-sized respectively. In vehicle experiments, the car clustering routine gave a true classification precision rate of 79\% for car-sized objects. Occasional errors occurred from clustering shrubbery or buildings as large vehicles. In total, 36\% of objects correctly identified as car-sized were not in fact cars; the true classification rate for cars was 51\%. Person clustering routine gave a true classification rate for person-sized objects of 89\%. Signs, fence posts, and telephone poles were the most common sources of errant person clusters because their lidar signature is similar to that of a human; false positives accounted for 36\% of the person-sized detections. The true classification rate of actual persons was 55\%. For the purpose of demonstrating joint perception solution in varying weather conditions, a simple robust clustering method was selected over high-fidelity, computationally complex, and brittle methods. Achieving 95\% confidence of object classification using the presented clustering routine's true classification rates of 51\% for cars and 55\% for pedestrians with the Ibeo Lidar units reporting at 12.5 Hz requires tracking uniformly classified clusters for 6 seconds to classify a car or 1.2 seconds to classify a person.

Raw object detection rates for the uncontrolled experiment are presented in Table~\ref{tab:Detection-Rate}. Sensors have different FOV, range, mounting location perspective, recall, and precision rates which all jointly contribute to the differences in raw detection rate. In general, increased resolution increases object detection rate; the 8k camera most closely matches the lidar coverage in front of Skynet. Object tracking and classification average rates for the various sensor configurations coupled with radar are reported in Table~\ref{tab:Tracking-Rate}.  The most interesting conclusion of Table~\ref{tab:Tracking-Rate} is the similarity in object tracking rates between the C-8k+R and L+R which implies that real-time high resolution imagery detection could provide a reliable fair-weather alternative to lidar-based tracking for environments with well-spaced objects.  For low resolution camera runs, the car and person classification rates are significantly below the object tracking rates because radar, which has no classification information, is providing the vast majority of object measurements. Lidar also has low classification rates but higher overall object tracking rate due to its long sensor range and high recall rate. High resolution imagery provides both increased object tracking rates and increased classification rates as the camera has good recall and excellent classification precision. As expected, coupling 8k imagery with lidar and radar provides the highest overall object detection and classification rates for tracked objects; interestingly the camera and lidar provide complementary information when fused. Later in the paper, sensor performance is evaluated for adverse weather and quantitative tracking positional accuracy, both important for closely-spaced crowded environments. For quantitative evaluations, 2k imagery resolution and a front semi-circular sensor mask of 20m was used in order to minimize the differences between lidar and camera FOV and range.

Results from experiment trials evaluated with a complete sensor set C+L+R, and reduced sensor sets C+R, or L+R, across all four weather condition categories are summarized in Tables~\ref{tab:results_person}-\ref{tab:results_suburban}. Radar is included in both camera and lidar sensor set evaluations to improve estimation speed and robustness, and because, most road-worthy AGVs contain radar. The radar coverage on the front of the vehicle is very sparse and cannot detect pedestrians so radar only-tracking was not performed. 

\begin{table}[htbp]
  \centering
  \caption{Pedestrian Tracking Performance.}
     \resizebox{\linewidth}{!}{\begin{tabular}{| m{2cm} | >{\centering\arraybackslash}m{1.25cm} | >{\centering\arraybackslash}m{1.8cm} >{\centering\arraybackslash}m{2.0cm} | >{\centering\arraybackslash}m{2.2cm} >{\centering\arraybackslash}m{2.2cm} | >{\centering\arraybackslash}m{2cm} |}
    \toprule
    {\textbf{Sensors Used}} & {\textbf{Object Tracked}} & {\textbf{Range \mbox{Error} (m)}} & {\textbf{Bearing Error (rad)}} & {\textbf{Correct \mbox{Classification}}} & {\textbf{Mis-Classification}} & {\textbf{Number of Returns}} \\
    \midrule\midrule
    {C + L + R} & {0.995} & {1.230} & {0.081} & {0.864} & {0.018} & {3964} \\
    {C + R} & {0.638} & {1.482} & {0.169} & {0.962} & {0.003} & {373} \\
    {L + R} & {0.995} & {1.252} & {0.084} & {0.852} & {0.024} & {3972} \\
    \bottomrule
    \end{tabular}}%
  \label{tab:results_person}%
\end{table}%
\begin{table}[htbp]
  \centering
  \caption{Vehicle Tracking Performance}
     \resizebox{\linewidth}{!}{\begin{tabular}{| m{2cm} | >{\centering\arraybackslash}m{1.25cm} | >{\centering\arraybackslash}m{1.8cm} >{\centering\arraybackslash}m{2.0cm} | >{\centering\arraybackslash}m{2.2cm} >{\centering\arraybackslash}m{2.2cm} | >{\centering\arraybackslash}m{2cm} |}
    \toprule
    {\textbf{Sensors Used}} & {\textbf{Object Tracked}} & {\textbf{Range \mbox{Error} (m)}} & {\textbf{Bearing Error (rad)}} & {\textbf{Correct \mbox{Classification}}} & {\textbf{Mis-Classification}} & {\textbf{Number of Returns}} \\
    \midrule\midrule
    {C + L + R} & {0.986} & {1.477} & {0.058} & {0.811} & {0.000} & {2594} \\
    {C/H + R} & {0.655} & {1.578} & {0.110} & {0.818} & {0.000} & {1369} \\
    {C + R} & {0.818} & {1.531} & {0.102} & {0.875} & {0.000} & {1776} \\
    {L + R} & {0.982} & {1.410} & {0.053} & {0.509} & {0.000} & {2565} \\
    \bottomrule
    \end{tabular}}%
  \label{tab:results_suburban}%
\end{table}%

Table~\ref{tab:results_suburban} also includes the case where heading detections are directly measured from the camera, labeled as C/H+R. In C/H+R, heading non-Gaussian measurement ambiguity was poorly captured with a large Gaussian covariance on the measurement. Camera detections are also evaluated using the multiple hypothesis heading direction classifier shown in Equation~(\ref{eq:heading_class}) which split the heading measurement as a continuous angle to the vehicle length axis and discrete direction along the line; the heading multiple hypothesis method is labeled C+R. As shown in Table~\ref{tab:results_suburban}, having an accurate measure of the angle to the vehicle length axis proved useful and more accurate: C+R had more returns, more objects tracked, lower range and bearing errors and better classification rates than C/H+R. Properly orienting the vehicle heading significantly improved association and object state estimate for wheeled dynamics. All other analyses involving C+R in this paper utilized the multiple hypothesis heading measurement split method.

The downsampled camera sensor images reduced the number of car detections, resulting in a lower object tracking time fraction in C+R compared to L+R. The range and bearing errors were also larger for the C+R case compared to L+R. The full resolution image processing could improve the object tracking time fraction. The joint classification results from Tables~\ref{tab:results_person}-\ref{tab:results_suburban} show that the C+R case has significantly better classification performance for cars compared to L+R, but for pedestrians, the improvement is less pronounced. This trend corresponds with expectation that the raw sensor classification accuracy difference between camera and lidar is larger for cars than pedestrians. The combined performance of the C+L+R case across both pedestrians and cars shows a combination of decreased range and bearing errors compared to C+R and increased classification performance compared to L+R. Intuitively, the benefits of having all three sensors is clearly shown in the C+L+R case: lidar is excellent at detecting objects and metrical information, whereas the camera is excellent at classification. The joint fusion of all sensors achieves a much more accurate and robust solution.

\subsection{Controlled Experiments: Weather Conditions in Joint Data Association, Tracking, and Classification}

The car and pedestrian tracking results are combined to analyze the performance of the three sensor combinations across each weather condition category. Table~\ref{tab:results_lidar} contains the combined L+R performance. All weather conditions had similar object tracking rates. Classification rates were highest for Night and second best during Cloudy \& Wet conditions. Darker conditions from night and to a lesser extent from clouds provided reduced solar radiation noise for which the sensor had to contend. Surprisingly, the Cloudy \& Wet conditions provided better lidar performance than daytime and best overall range estimates. Wet objects tend to scatter lidar returns; one plausible explanation is that the dry Suburban and dry pedestrians provided improved reflected signal returns compared to the reduced background noise returns. In summary, precipitation conditions degraded lidar performance most drastically in classification but also in range and bearing, while lidar performance improved in darker conditions because there was less reflected solar radiation to interfere with the lidar.

\begin{table}[htbp]
  \centering
  \caption{Lidar + Radar (L+R) Performance in Weather}
     \resizebox{\linewidth}{!}{\begin{tabular}{| m{2.2cm} | >{\centering\arraybackslash}m{1.25cm} | >{\centering\arraybackslash}m{1.8cm} >{\centering\arraybackslash}m{2.0cm} | >{\centering\arraybackslash}m{2.2cm} >{\centering\arraybackslash}m{2.2cm} | >{\centering\arraybackslash}m{2cm} |}
    \toprule
    {\textbf{Description}} & {\textbf{Object Tracked}} & {\textbf{Range \mbox{Error} (m)}} & {\textbf{Bearing Error (rad)}} & {\textbf{Correct \mbox{Classification}}} & {\textbf{Mis-Classification}} & {\textbf{Number of Returns}} \\
    \midrule\midrule
    {Cloudy \& Wet} & {0.988} & {0.700} & {0.063} & {0.726} & {0.000} & {1493} \\
    {Sunny} & {0.999} & {1.437} & {0.083} & {0.664} & {0.000} & {1480} \\
    {Night} & {0.987} & {1.424} & {0.060} & {0.872} & {0.000} & {1781} \\
    {Snow \& Rain} & {0.986} & {1.484} & {0.085} & {0.601} & {0.052} & {1783} \\
    \bottomrule
    \end{tabular}}%
  \label{tab:results_lidar}%
\end{table}%

Table~\ref{tab:results_camera} contains the combined C+R performance. Similar to results shown in Tables~\ref{tab:results_person}-\ref{tab:results_suburban}, C+R object track rates in individual weather categories were all lower than the L+R weather categories because the downsampled camera sensor images reduced the number of car detections. Night had the worst object tracking fraction due to poor lighting conditions. Cloudy \& Wet conditions provided best tracking fraction due to the uniform diffuse lighting conditions. Sunny daytime conditions have more glare and stark shadows to contend with than Cloudy conditions, resulting in worst range and bearing errors. One unexpected result was Night range and bearing error were less than Sunny conditions; it is hypothesized that this is due to shadow ambiguities in estimating an object bounding box. As expected, Snow \& Rain precipitation degraded the object tracking fraction and also degraded range and bearing estimates. Surprisingly, classification rates were similar for Cloudy \& Wet, Sunny, and Snow \& Rain. The degraded Night classification, due to poor lighting conditions, was worse than Night classification in the L+R case. In summary, bad lighting conditions, especially at night, were more detrimental to camera performance than weather conditions.

\begin{table}[htbp]
  \centering
  \caption{Camera + Radar (C+R) Performance in Weather}
     \resizebox{\linewidth}{!}{\begin{tabular}{| m{2.2cm} | >{\centering\arraybackslash}m{1.25cm} | >{\centering\arraybackslash}m{1.8cm} >{\centering\arraybackslash}m{2.0cm} | >{\centering\arraybackslash}m{2.2cm} >{\centering\arraybackslash}m{2.2cm} | >{\centering\arraybackslash}m{2cm} |}
    \toprule
    {\textbf{Description}} & {\textbf{Object Tracked}} & {\textbf{Range \mbox{Error} (m)}} & {\textbf{Bearing Error (rad)}} & {\textbf{Correct \mbox{Classification}}} & {\textbf{Mis-Classification}} & {\textbf{Number of Returns}} \\
    \midrule\midrule
    {Cloudy \& Wet} & {0.916} & {1.414} & {0.083} & {0.971} & {0.000} & {552} \\
    {Sunny} & {0.867} & {1.657} & {0.121} & {0.971} & {0.002} & {588} \\
    {Night} & {0.620} & {1.425} & {0.112} & {0.590} & {0.000} & {432} \\
    {Snow \& Rain} & {0.739} & {1.627} & {0.127} & {0.951} & {0.000} & {450} \\
    \bottomrule
    \end{tabular}}%
  \label{tab:results_camera}%
\end{table}%

Table~\ref{tab:results_full} contains the complete sensor set C+L+R performance. Across all weather conditions the tracked object fraction was \~99\%. Range and bearing errors are less than the C+R sensor set and correct classification is improved over the L+R set which is consistent with the controlled experiment results of Tables~\ref{tab:results_person}-\ref{tab:results_suburban}. As in the L+R sensor set, the C+L+R Cloudy \& Wet conditions provided the best overall range estimates. The number of returns for L+R is three to four times that of C+R. The addition of camera data improved classification rates across all weather scenarios in C+L+R compared to L+R. However, due to the lower total number of camera returns, there were many sections tracked by L+R for which no camera information was available; thus the C+L+R classification rates are lower than that of C+R because of a discrepancy in the number of returns. Given a similar number of C+R and L+R returns, it is expected that the C+L+R classification rate would match or exceed that of the C+R. The mis-classification rate is nearly zero for all presented examples; the probabilistic filter is combining raw sensor classification information in an unbiased manner.

\begin{table}[htbp]
  \centering
  \caption{Full Sensor Set (C+L+R) Performance in Weather}
    \resizebox{\linewidth}{!}{\begin{tabular}{| m{2.2cm} | >{\centering\arraybackslash}m{1.25cm} | >{\centering\arraybackslash}m{1.8cm} >{\centering\arraybackslash}m{2.0cm} | >{\centering\arraybackslash}m{2.2cm} >{\centering\arraybackslash}m{2.2cm} | >{\centering\arraybackslash}m{2cm} |}
    \toprule
    {\textbf{Description}} & {\textbf{Object Tracked}} & {\textbf{Range \mbox{Error} (m)}} & {\textbf{Bearing \mbox{Error} (rad)}} & {\textbf{Correct \mbox{Classification}}} & {\textbf{Mis-Classification}} & {\textbf{Number of Returns}} \\
    \midrule\midrule
    {\mbox{Cloudy \&} Wet} & {0.985} & {0.825} & {0.056} & {0.876} & {0.000} & {1517} \\
    {Sunny} & {0.998} & {1.544} & {0.077} & {0.875} & {0.000} & {1455} \\
    {Night} & {0.989} & {1.320} & {0.070} & {0.889} & {0.000} & {1779} \\
    {Snow \& Rain} & {0.995} & {1.499} & {0.085} & {0.744} & {0.040} & {1807} \\
    \bottomrule
    \end{tabular}}%
  \label{tab:results_full}%
\end{table}%

The presented results demonstrate performance of a state-of-the-art camera detector and lidar configuration supplemented with radar exhibiting reliable performance across varying weather scenarios. Sensor diversity and a probabilistic filter are critical for adding robustness to performance across weather scenarios.

\subsection{Urban Driving Experiments: Qualitative Discussion of Performance in Weather Conditions}

Experiments were conducted through CollegeTown near Cornell's campus and downtown Ithaca to study performance in real-world scenarios involving diverse vehicle and pedestrian traffic on typical busy streets in various weather scenarios. Skynet was driven in each weather condition, Snow, Rain, Sunny, Cloudy, Wet \& Humid, and Night, for over 30 miles and 2 hours of time, at speeds up to 35 miles per hour, in an assortment of urban conditions including two-lane one-way roads, two-lane two-way roads, assorted intersections involving pedestrians, and vehicles controlled with lights and signs, and around a number of building and neighborhood styles including downtown businesses, housing, and commercial box-store areas.

Observations made about overall sensor performance in clear weather are as follows: radar has few false positives, lidar is highly accurate at depth measurements, camera is highly accurate at correctly classifying vehicle and pedestrian detections, albeit with dependency on processing resolution; camera detection range is also dependent on processing resolution.

Active and passive air-borne disturbances are present during and after precipitation, such as rain or snow and during heavy fog. Water droplets and snow flakes could be detected in lidar and were visible in the camera frame. As has been published elsewhere \cite{Mercedes-S-Class}, \cite{agv-wired}, precipitation generally did not affect radars in a noticeable way in the experiments. The Ibeo XT lidar units have on-board rain drop filtering that examines the intensity and time response of individual lidar returns; if a double return is detected and the first response is lower in amplitude or shorter in duration, the first return is assumed to be a rain drop and the second return is the actual object. In principle, filtering airborne disturbances at the sensor's receiver processor could be beneficial, but in practice only a small portion of the rain returns were negated and had little effect on filtering snow flake returns.

\begin{figure} [h!]
    \centering
    \includegraphics[width=2.5in]{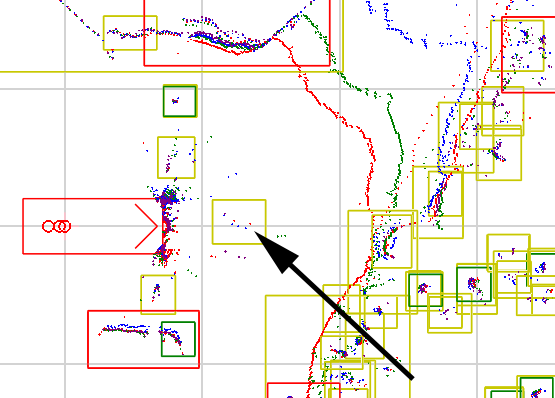}
    \includegraphics[width=2.5in]{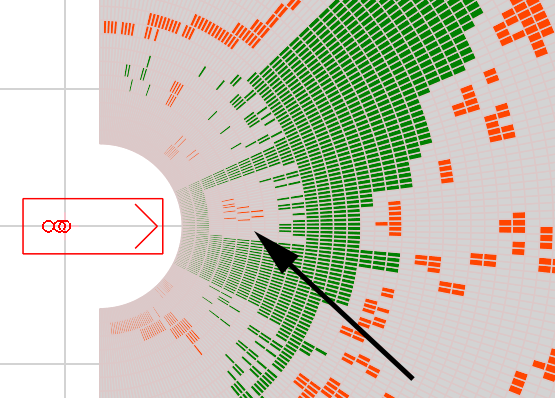}\\
    \includegraphics[width=5.0in]{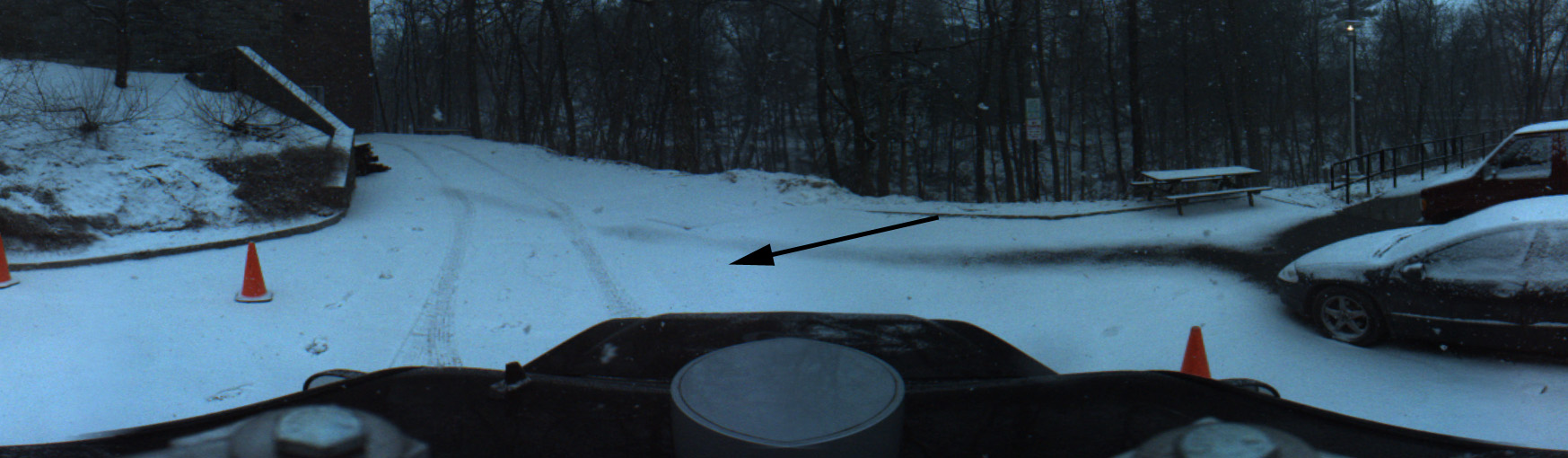}
    \caption{Experimental time snapshot showing lidar returns from snow flakes. In top-down depictions, Skynet is facing right. Upper left: lidar and clustering. Upper right: occupancy grid. Bottom: view of scene. The black arrow point to the same location, and denote lidar sensor returns and the resulting occupancy in the occupancy grid due to detecting snow flakes as objects.}
    \label{fig:occupancy-grid}
\end{figure}

The occupancy grid used for safety and collision avoidance was also susceptible to weather conditions. As shown in Figure~\ref{fig:occupancy-grid}, the lidar can return measurements of snow which are in turn passed to the occupancy grid. Snowflakes white color, larger size, and slower dynamics can cause returns of multiple lidar beams. Unfortunately, these returns are indistinguishable from object returns in intensity, and, importantly, they are close in proximity to the vehicle, creating safety concerns. Instead of lidar, commercial cars use sonar and radar for object detection and avoidance because they are less susceptible to airborne precipitation. When considering all-weather driving, heavy reliance on lidar in AGV research could present problems to commercialization.

\begin{figure} [h]
    \centering
    \includegraphics[width=4.0in]{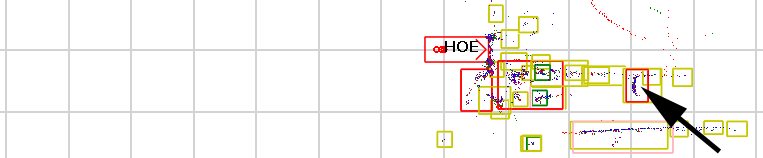}
    \includegraphics[width=4.0in]{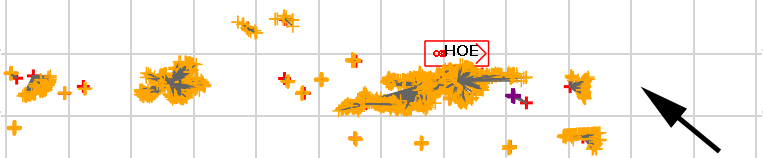}\\
    \includegraphics[width=6.5in]{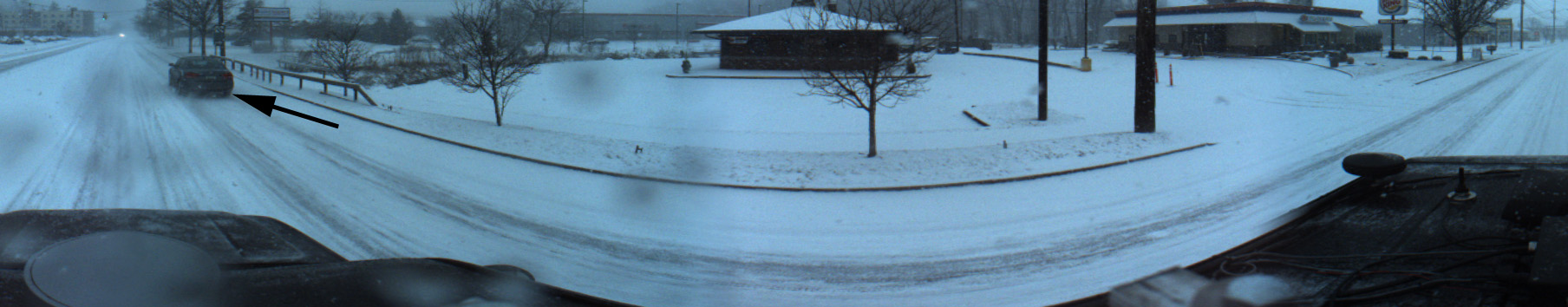}
    \caption{Lidar returns of snow trailing behind vehicle causes errant clustering and therefore errant tracked objects. Arrows point to a vehicle object location.  Top: lidar returns and clustering. Middle: tracked objects. Bottom: Skynet faces left and shows clear road.}
    \label{fig:snow-tracking}
\end{figure}

Water and snow blown behind other vehicles in front of Skynet also made it difficult to track objects with lidar. Figure~\ref{fig:snow-tracking} shows a snow example where dozens of phantom objects are being tracked when the car is clearly visible in the camera's frame of view. Radar was completely unaffected by the water and snow spray, while the camera was only affected when the density of spray was strong enough to significantly cloud the lens and hide the vehicle. In the lidar returns, the trailing spray was clustered arbitrarily as person- or car-sized depending on size of spray pattern and caused the RBPF to birth phantom tracked objects which followed behind the car. Dozens of phantom objects occupied several car-lengths of space, trailing the car while the actual car object was not reliably tracked or classified. In Figure~\ref{fig:snow-tracking}, the black arrow points to the true location of the car, but no tracked object is estimated in the car's true location.

\begin{figure} [h!]
    \centering
    \includegraphics[width=2.5in]{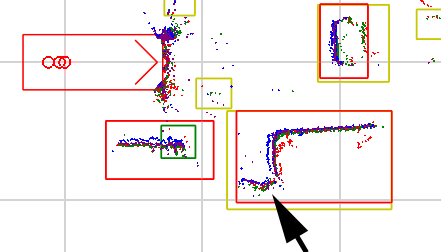}
    \includegraphics[width=2.5in]{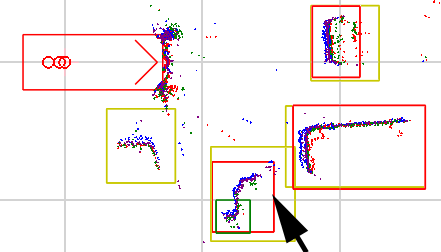}\\
    \includegraphics[width=5.0in]{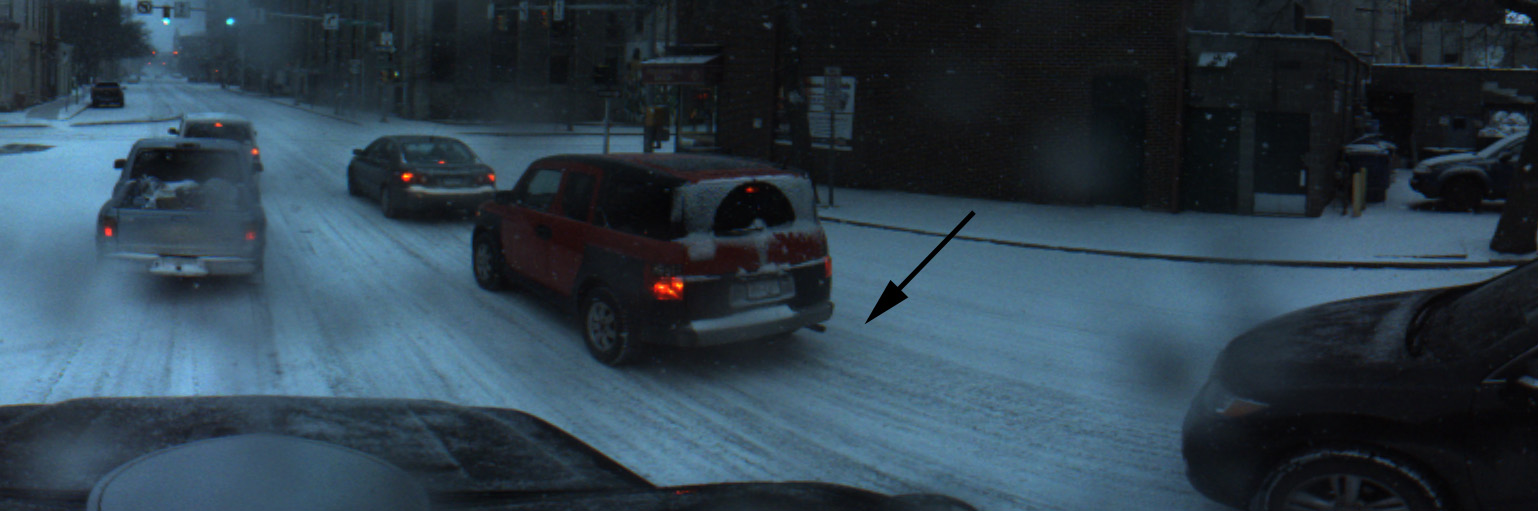}
    \caption{Example of an exhaust plume causing lidar sensor returns. Upper left: lidar and clustering. Right: same scene 2 seconds later as vehicle moves through green light. Bottom: view of scene. Black arrow points to example of optically invisible plume being detected by lidar and clustered as a person (green box) and car (red box).}
    \label{fig:exhaust-plume}
\end{figure}

The most surprising airborne phenomenon was the observation that exhaust plumes can cause lidar returns. During cool conditions, vehicle exhaust or steam venting from city street tunnels into ambient temperatures with a relatively high dew-point caused the exhaust to condense into clouds and create large plumes that created lidar returns. Interestingly, exhaust and steam plumes that are barely visible to the human eye or cameras in the optical spectrum can cause significant lidar returns. Figure~\ref{fig:exhaust-plume} shows an example of lidar returns at two time instances and the corresponding camera image; black arrows point to the exhaust plume location. Initially, the plume was clustered with the vehicle, but as the vehicle drove through the green light, the plume was blown behind the vehicle and clustered as both a person-sized and car-sized object. This phenomenon was most prevalent when driving in winter, cool, or rainy conditions and largely non-existent when ambient temperatures were warm and the dew-point was low.

Sensor fouling was most common in snow and rain, but can also occur in dusty conditions. Wipers, such as those typically found on windshields, are important for all sensors; given enough sensor surface accumulation of snow, ice, or dirt, radar, camera, and lidar sensors malfunction. Radar is robust to water surface accumulation on the sensor but snow pack and ice accumulation from highway driving can disable radar. Camera and lidar sensors are both sensitive to any sensor surface accumulation that blocks light; the lidar could operate with a wet lens cover; the camera scenes of pedestrians and cars through a wet lens became unrecognizable to the detector. The Velodyne lidar naturally stays clean, given its rotation which limits water and snow from directly hitting the lens. What water does hit the lens tends to blow off from wind and centrifugal force.

\begin{figure} [h]
    \centering
    \includegraphics[height=1.5in]{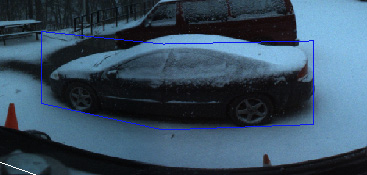}
    \includegraphics[height=1.5in]{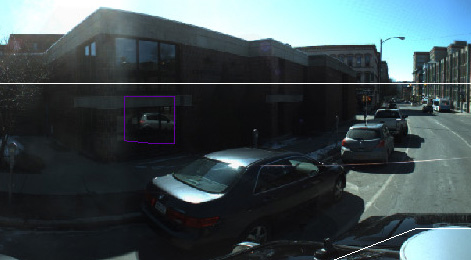}
    \caption{Left: detection of car in snow. Right: detection of car in store window reflection.}
    \label{fig:camera-reflection}
\end{figure}

Environmental surface accumulation has the potential to cause a variety of unexpected sensor behaviors. For lidar, wet object surfaces decreased return rates while snowy object surfaces increased return rates. The camera was unaffected in general by light environmental surface accumulation, but under heavy snow the camera eventually was unable to distinguish edges of objects and their environment. Figure~\ref{fig:camera-reflection} shows an example camera detection of a snow covered car.

Reflections were noticed in the camera when driving by shiny buildings, as captured in Figure~\ref{fig:camera-reflection}. This example poses less of a safety issue in that its validity could be reasoned about using the detected location of the obstacle. Selecting an exposure setting for the cameras was somewhat challenging. For simplicity, a uniform exposure across lenses was selected to support easy panorama creation for detection processing. However, uniform exposure creates problems for areas where lighting has gross variations, such as dusk, dawn, and oncoming headlights. These lighting problems are magnified with a wet road surface due to reflections.

It has been reported \cite{Nvidia-radar-ex} that radar can have problems with shiny and reflective glass or objects, but this was not observed in Skynet recorded data.

\begin{figure} [h]
    \centering
    \includegraphics[height=1.5in]{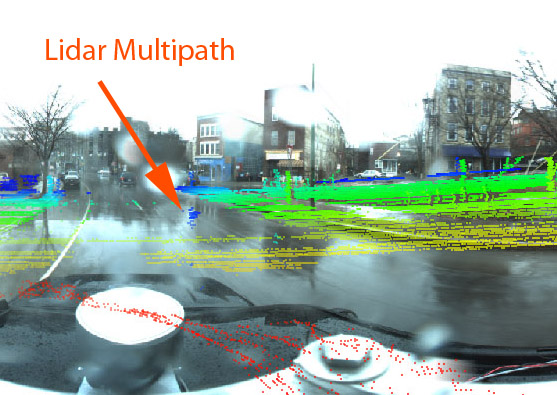}
    \includegraphics[height=1.5in]{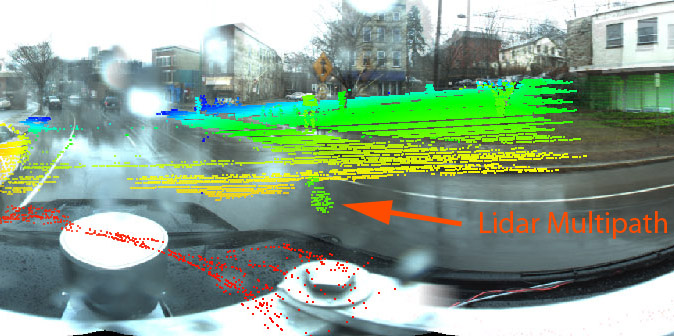}
    \includegraphics[height=1.5in]{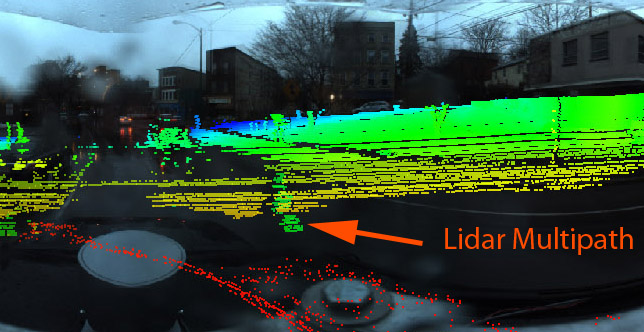}
    \caption{Examples of lidar multipath from wet conditions.}
    \label{fig:lidar-multipath}
\end{figure}

Multi-path lidar was a significant problem on wet surfaces, as shown in the examples in Figure~\ref{fig:lidar-multipath}. In airborne topology mapping, \cite{lidar-forest} multi-path lidar has been reported and is typically avoided by controlling the inclination angle to ground targets to be nearly vertical. However, smooth wet roads cause lidar to reflect on its way to or from the target. The lidar collector which receives the return, determines the angle corresponding to the range measurement. If the lidar beam initially reflected off the ground, and then reflected off an object before returning to the collector, the object is projected to be farther away than it actually is due to the longer round trip path. Over-estimating the distance to a target can be dangerous and lead to AGV collisions. Measurement gating might offer some potential ways to alleviate  the multi-path problem.

\begin{figure} [h]
    \centering
    \includegraphics[width=3.2in]{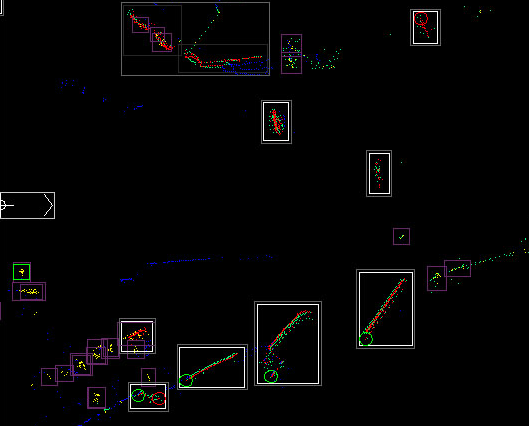}
    \includegraphics[width=3.2in]{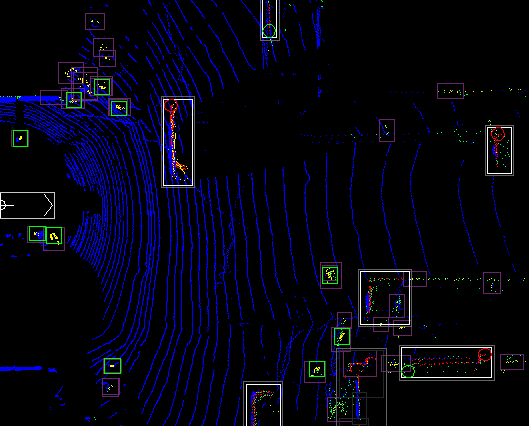}
    \caption{Ground view of lidar returns inside the lidar clustering routine. Blue dots are classified as ground plane; white boxes are car-sized objects; green boxes are person-sized objects. Left: Ground plane estimated below actual ground level from sparse multi-path returns in wet conditions. Right: Typical ground plane estimate in dry conditions.}
    \label{fig:ground-plane}
\end{figure}

If the lidar beam first reflects off the object, and then reflects off the ground on the return path, the range to the object is projected to go through the ground. Many AGVs also estimate the location of the ground or roadways with some form of ground plane detection; examples of multi-path lidar returns projected to have originated from below ground level, as defined by Skynet's ground plane detection, are shown in Figure~\ref{fig:ground-plane}. Due to lidar scattering off the wet road surface creating multi-path returns, the ground plane was estimated below the actual ground level shown in Figure~\ref{fig:ground-plane} left. For comparison, Figure~\ref{fig:ground-plane} right shows an example of a typical ground plane estimate in dry conditions with dense lidar returns. Low ground plane estimation could lead to over-estimating object range calculations for camera detections, which could lead to AGV collisions.

From the above experiments, in both fair weather and not, camera, lidar, and radar sensors were found to compliment each other. Lidar showed improved returns from snow covered objects and excellent night performance. However, lidar challenges included multi-path reflections on rainy roads, difficulty with ground plane estimation in wet conditions, clustering and tracking issues with snow including that blown up by other traffic or wind, response degradation to wet vehicles, response to exhaust plumes in cold and wet conditions, and errant occupancy grid response. The camera was more robust to precipitation, and provided the most accurate classifications, but struggled with dark conditions or lighting variations within a single scene. Radar was most robust to different weather conditions and  provides accurate velocity information, but typically cannot detect pedestrians and lacks some of the depth, shape, and size accuracy of lidar or the classification accuracy of the camera. Occasional clutter also obscures radar returns for distant objects or closely spaced objects at similar distances and speeds.

\section{Conclusion}

A novel real-time probabilistic joint data association, tracking, and classification system for an autonomous ground vehicle is formulated. Additionally, a state-of-the-art vision detection algorithm that includes heading detection for autonomous ground vehicle applications is integrated and compared. With the incorporation of lidar clustering, radar sensors, and pose, a real-time demonstration of the joint probabilistic perception algorithm was conducted in varying weather conditions and using different subsets of sensors. Monte Carlo simulations, repeatable controlled experiments, and a lengthy real-world urban data collection demonstrated performance and identified new challenges with weather perception and unique capabilities of a joint association, tracking, and classification solution.

Many observations were made regarding autonomous ground vehicle performance in weather. In general, lidar was most brittle to laser blockages, multi-path returns, airborne precipitation, and wet surfaces, and most robust to lighting conditions, while providing object shape and size information. The camera was most brittle to dark lighting conditions and glare, but was more robust to precipitation than lidar. Glare, which was often present in wet conditions from headlights or sunshine, reflected brightly off road, vehicle, and building surfaces, making camera exposure selection difficult and degrading object detection performance. Radar has the best robustness in performance to all weather conditions, but often cannot detect pedestrians and provides less information about object shape, size, and classification than lidar or camera.

Given the various limitations of existing sensors, there is much opportunity for future development of sensor hardware, sensor data processing, and perception algorithm advancements. Cheap and reliable lens cleaning for cameras and to a lesser degree for lidar and radar are necessary for reliable operation in any form of precipitation. Improving the dynamic range of cameras, composing high dynamic range images, or actively modulating exposure across the CCD may provide some potential for improving camera operation at night and during high glare conditions. Extending tracking of precipitation \cite{dlp-headlights} with classification to categorize various weather phenomenon could improve individual sensor performance or perception system performance. As computing power increases and image detection methods advance, higher resolution image processing could enable higher detection rates and longer detection ranges for vision. Direct estimation of the current weather condition could allow active of toggling sensors, falling back to a reduced base sensor set, or weighting sensor returns if each sensor's performance is accurately characterized across weather conditions; for example, spurious lidar returns of snow, exhaust plumes, or phantom occupancy grid objects could be automatically discarded if the weather condition and sensor's weather sensitivities were known by the vehicle. Ground plane estimates could be used to reason and correct for multi-path lidar returns using hypothesis gating. Joint sensor validation could enable discarding of bad weather returns, for example the camera could help discard exhaust plumes and blowing snow lidar returns. In summary, sensor diversity and joint estimation of data association, tracking, and classification proved beneficial towards robust performance in all-weather conditions and provides a framework for future advancements.

\subsubsection*{Acknowledgments}
The authors would like to thank Trimble for providing Omnnistar HP DGPS corrections service, NVIDIA for providing a GTX980 GPU, Kevin Wyffels for insightful discussions on the classification formulation, and You Won Park for constructing the Ladybug3 Camera Mount.

\bibliography{Weather_Perception_AGV}

\begin{thebibliography}{}

\bibitem[Ackerman, 2014]{KAIST-weather}
Ackerman, E. (2014).
\newblock {Korean Competition Shows Weather Still a Challenge for Autonomous
  Cars}.
\newblock In {\em {IEEE Spectrum, Cars that Think Website}}.
\newblock
  http://spectrum.ieee.org/cars-that-think/transportation/advanced-cars/japan-competition-shows-weather-still-a-challenge-for-autonomous-cars.

\bibitem[Angelova et~al., 2015]{Angelova-2015}
Angelova, A., Krizhevsky, A., and Vanhoucke, V. (2015).
\newblock {Pedestrian detection with a large-field-of-view deep network}.
\newblock In {\em {Robotics and Automation (ICRA), 2015 IEEE International
  Conference on}}, pages 704--711.

\bibitem[Bar-Shalom et~al., 2001]{Bar-Shalom}
Bar-Shalom, Y., Li, X.~R., and Kirubarajan, T. (2001).
\newblock {\em {Estimation with Applications to Tracking and Navigation}}.
\newblock {John Wiley \& Sons, Inc.}

\bibitem[Cai et~al., 2015]{Vasconcelos-Pedestrian}
Cai, Z., Saberian, M.~J., and Vasconcelos, N. (2015).
\newblock Learning complexity-aware cascades for deep pedestrian detection.
\newblock {\em CoRR}, abs/1507.05348.

\bibitem[Dalal and Triggs, 2005]{HOG-detections}
Dalal, N. and Triggs, B. (2005).
\newblock Histograms of oriented gradients for human detection.
\newblock In {\em Computer Vision and Pattern Recognition (CVPR), IEEE Computer
  Society Conference on}, volume~1, pages 886--893 vol. 1.

\bibitem[Deng, 2014]{deep-learning-tutorial}
Deng, L. (2014).
\newblock {A Tutorial Survey of Architectures, Algorithms, and Applications for
  Deep Learning}.
\newblock {\em {APSIPA Transactions on Signal and Information Processing}}.
\newblock http://research.microsoft.com/apps/pubs/default.aspx?id=204048.

\bibitem[Dickmann et~al., 2014]{Mercedes-S-Class}
Dickmann, J., Appenrodt, N., and Brenk, C. (2014).
\newblock {How We Gave Sight to the Mercedes Robotic Car}.
\newblock In {\em {IEEE Spectrum, Transportation Website}}.
\newblock
  http://spectrum.ieee.org/transportation/self-driving/how-we-gave-sight-to-the-mercedes-robotic-car.

\bibitem[Douillard et~al., 2011]{Douillard-ICRA2011}
Douillard, B., Underwood, J., Kuntz, N., Vlaskine, V., Quadros, A., Morton, P.,
  and Frenkel, A. (2011).
\newblock {On the segmentation of 3D LIDAR point clouds}.
\newblock In {\em {Robotics and Automation (ICRA), 2011 IEEE International
  Conference on}}, pages 2798--2805.

\bibitem[Everingham et~al., 2010]{everingham-PASCAL10}
Everingham, M., Van~Gool, L., Williams, C. K.~I., Winn, J., and Zisserman, A.
  ({2010}).
\newblock {The Pascal Visual Object Classes ({VOC}) Challenge}.
\newblock {\em {International Journal of Computer Vision (IJCV)}}, {88}({2}).

\bibitem[Felzenszwalb et~al., 2010a]{DPM-TPAMI}
Felzenszwalb, P., Girshick, R., McAllester, D., and Ramanan, D. (2010a).
\newblock Object detection with discriminatively trained part-based models.
\newblock {\em Pattern Analysis and Machine Intelligence, IEEE Transactions
  on}, 32(9):1627--1645.

\bibitem[Felzenszwalb et~al., 2010b]{Felzenszwalb-DPM}
Felzenszwalb, P.~F., Girshick, R.~B., McAllester, D., and Ramanan, D. (2010b).
\newblock {Object Detection with Discriminatively Trained Part Based Models}.
\newblock {\em {IEEE Transactions on Pattern Analysis and Machine
  Intelligence}}, 32(9):1627--1645.

\bibitem[Freund and Schapire, 1999]{Adaboost}
Freund, Y. and Schapire, R.~E. (1999).
\newblock {A Short Introduction to Boosting}.
\newblock {\em {Journal of Japanese Society for Artificial Intelligence}},
  14(5):771--780.

\bibitem[Gatziolis and Andersen, 2008]{lidar-forest}
Gatziolis, D. and Andersen, H.-E. (2008).
\newblock {A Guide to LIDAR Data Acquisition and Processing for the Forests of
  the Pacific Northwest}.
\newblock Technical Report PNW-GTR-768, United States Department of
  Agriculture: Forest Service Pacific Northwest Research Station.

\bibitem[Geiger et~al., 2013]{Geiger2013IJRR}
Geiger, A., Lenz, P., Stiller, C., and Urtasun, R. (2013).
\newblock {Vision meets Robotics: The KITTI Dataset}.
\newblock {\em {International Journal of Robotics Research (IJRR)}}.

\bibitem[Girshick, 2015]{Girshick-ICCV-2015}
Girshick, R. (2015).
\newblock {Fast R-CNN}.
\newblock In {\em IEEE International Conference on Computer Vision (ICCV)}.

\bibitem[Girshick et~al., 2014]{Girshick-CVPR-2014}
Girshick, R., Donahue, J., Darrell, T., and Malik, J. (2014).
\newblock {Rich Feature Hierarchies for Accurate Object Detection and Semantic
  Segmentation}.
\newblock In {\em Computer Vision and Pattern Recognition (CVPR), IEEE
  Conference on}.

\bibitem[Girshick et~al., 2012]{voc-release5}
Girshick, R.~B., Felzenszwalb, P.~F., and McAllester, D. (2012).
\newblock {Discriminatively Trained Deformable Part Models, Release 5}.
\newblock {http://people.cs.uchicago.edu/~rbg/latent-release5/}.

\bibitem[He et~al., 2015]{He-arxiv-2015}
He, K., Zhang, X., Ren, S., and Sun, J. (2015).
\newblock {Deep Residual Learning for Image Recognition}.
\newblock {http://arxiv.org/pdf/1512.03385v1.pdf}.

\bibitem[Held et~al., 2012]{Held-ICRA2012}
Held, D., Levinson, J., and Thrun, S. (2012).
\newblock {A probabilistic framework for car detection in images using context
  and scale}.
\newblock In {\em {Robotics and Automation (ICRA), 2012 IEEE International
  Conference on}}, pages 1628--1634.

\bibitem[Held et~al., 2013]{Held-ICRA2013}
Held, D., Levinson, J., and Thrun, S. (2013).
\newblock {Precision tracking with sparse 3D and dense color 2D data}.
\newblock In {\em {Robotics and Automation (ICRA), 2013 IEEE International
  Conference on}}, pages 1138--1145.

\bibitem[Held et~al., 2014]{Held-RSS-14}
Held, D., Levinson, J., Thrun, S., and Savarese, S. (2014).
\newblock {Combining 3D Shape, Color, and Motion for Robust Anytime Tracking}.
\newblock In {\em {Proceedings of Robotics: Science and Systems}}, Berkeley,
  USA.

\bibitem[Huang, 2015]{DrivePX}
Huang, J.~S. (2015).
\newblock {Drive PX}.
\newblock In {\em {Consumer Electronics Show}}.
\newblock http://www.nvidia.com/object/drive-px.html.

\bibitem[Ilg et~al., 2014]{Ilg-ICRA2014}
Ilg, E., Kuummerle, R., Burgard, W., and Brox, T. (2014).
\newblock {Reconstruction of rigid body models from motion distorted laser
  range data using optical flow}.
\newblock In {\em {Robotics and Automation (ICRA), 2014 IEEE International
  Conference on}}, pages 4627--4632.

\bibitem[Korchev et~al., 2013]{Korchev-RTLidar}
Korchev, D., Cheng, S., Owechko, Y., and Kim, K. (2013).
\newblock {On Real-Time LIDAR Data Segmentation and Classification}.
\newblock In {\em {Proceedings of Image Processing, Computer Vision, \& Pattern
  Recognition ICPV}}.

\bibitem[Krizhevsky et~al., 2012]{deep-learning}
Krizhevsky, A., Sutskever, I., and Hinton, G.~E. (2012).
\newblock {ImageNet Classification with Deep Convolutional Neural Networks}.
\newblock In {\em {NIPS: Neural Information Processing Systems}}, Lake Tahoe,
  Nevada.

\bibitem[Matzen and Snavely, 2013]{nyc3dcars}
Matzen, K. and Snavely, N. (2013).
\newblock {NYC3DCars: A Dataset of 3D Vehicles in Geographic Context}.
\newblock In {\em {Proc. Int. Conf. on Computer Vision}}.

\bibitem[Miller and Campbell, 2012]{pose-skynet}
Miller, I. and Campbell, M. (2012).
\newblock {Sensitivity Analysis of a Tightly-Coupled GPS/INS System for
  Autonomous Navigation}.
\newblock {\em {Aerospace and Electronic Systems, IEEE Transactions on}},
  48(2):1115--1135.

\bibitem[Miller et~al., 2011a]{RBPF-Skynet}
Miller, I., Campbell, M., and Huttenlocher, D. (2011a).
\newblock {Efficient Unbiased Tracking of Multiple Dynamic Obstacles Under
  Large Viewpoint Changes}.
\newblock {\em {Robotics, IEEE Transactions on}}, 27(1):29--46.

\bibitem[Miller et~al., 2011b]{posterior-pose}
Miller, I., Campbell, M., and Huttenlocher, D. (2011b).
\newblock {Map-aided localization in sparse global positioning system
  environments using vision and particle filtering}.
\newblock {\em {Journal of Field Robotics}}, 28(5):619--643.

\bibitem[Miller et~al., 2008]{skynet}
Miller, I., Campbell, M., Huttenlocher, D., Kline, F.-R., Nathan, A., Lupashin,
  S., Catlin, J., Schimpf, B., Moran, P., Zych, N., Garcia, E., Kurdziel, M.,
  and Fujishima, H. (2008).
\newblock {Team Cornell's Skynet: Robust perception and planning in an urban
  environment}.
\newblock {\em {Journal of Field Robotics}}, 25(8):493--527.

\bibitem[Monteiro et~al., 2006]{Montiero-IROS2006}
Monteiro, G., Premebida, C., Peixoto, P., and Nunes, U. (2006).
\newblock {Tracking and classification of dynamic obstacles using laser range
  finder and vision}.
\newblock In {\em {Proc. of the IEEE/RSJ International Conference on
  Intelligent Robots and Systems (IROS)}}.

\bibitem[{NHTSA}, 2014]{NHTSA-guidelines}
{NHTSA} (2014).
\newblock {\em {Preliminary Statement of Policy Concerning Automated
  Vehicles}}.
\newblock {National Highway Traffic Safety Administration}.

\bibitem[Ramachandra, 2000]{KF-Radar}
Ramachandra, K.~V. (2000).
\newblock {\em {Kalman Filtering Techniques for Radar Tracking}}.
\newblock {Marcel Dekker, Inc.}

\bibitem[Ren et~al., 2015]{NIPS-2015}
Ren, S., He, K., Girshick, R., and Sun, J. (2015).
\newblock {Faster R-CNN: Towards Real-Time Object Detection with Region
  Proposal Networks}.
\newblock In {\em {Neural Information Processing Systems (NIPS)}}. {Microsoft
  Research}.
\newblock http://arxiv.org/pdf/1506.01497v2.pdf.

\bibitem[Ristic et~al., 2004]{mm-kinematics-classification}
Ristic, B., Gordon, N., and Bessell, A. (2004).
\newblock {On target classification using kinematic data}.
\newblock {\em {Information Fusion}}, 5:15--21.

\bibitem[Ross, 2015]{Nvidia-radar-ex}
Ross, P.~E. (2015).
\newblock {Nvidia Wants to Build the Robocar's Brain}.
\newblock In {\em {IEEE Spectrum, Cars that Think Website}}.
\newblock
  http://spectrum.ieee.org/cars-that-think/transportation/self-driving/nvidia-wants-to-build-the-robocars-brain.

\bibitem[Sadeghi and Forsyth, 2014]{DPM-30Hz}
Sadeghi, M.~A. and Forsyth, D. (2014).
\newblock {30Hz Object Detection with DPM V5}.
\newblock In Fleet, D., Pajdla, T., Schiele, B., and Tuytelaars, T., editors,
  {\em {Computer Vision -- ECCV 2014}}, volume 8689 of {\em Lecture Notes in
  Computer Science}, pages 65--79. {Springer International Publishing}.

\bibitem[{SAE}, 2013]{SAE-guidelines}
{SAE} (2013).
\newblock {\em {Taxonomy and Definitions for Terms Related to On-Road Motor
  Vehicle Automated Driving Systems - J3016}}.
\newblock {Society of Automotive Engineers: On-Road Automated Vehicle Standards
  Committee}.

\bibitem[Schoenberg et~al., 2012]{GSF}
Schoenberg, J.~R., Campbell, M., and Miller, I. (2012).
\newblock {Posterior representation with a multi-modal likelihood using the
  gaussian sum filter for localization in a known map}.
\newblock {\em {Journal of Field Robotics}}, 29(2):240--257.

\bibitem[Tamburo et~al., 2014]{dlp-headlights}
Tamburo, R., Nurvitadhi, E., Chugh, A., Chen, M., Rowe, A., Kanade, T., and
  Narasimhan, S. (2014).
\newblock Programmable automotive headlights.
\newblock In {\em European Conference of Computer Vision (ECCV)}, volume 8692
  of {\em Lecture Notes in Computer Science}, pages 750--765. Springer
  International Publishing.

\bibitem[Teichman et~al., 2011]{Teichman-ICRA2013}
Teichman, A., Levinson, J., and Thrun, S. (2011).
\newblock {Towards 3D object recognition via classification of arbitrary object
  tracks}.
\newblock In {\em {Robotics and Automation (ICRA), 2011 IEEE International
  Conference on}}, pages 4034--4041.

\bibitem[Teichman and Thrun, 2011]{Teichman-ARSO2011}
Teichman, A. and Thrun, S. (2011).
\newblock {Practical object recognition in autonomous driving and beyond}.
\newblock In {\em {Advanced Robotics and its Social Impacts (ARSO), 2011 IEEE
  Workshop on}}, pages 35--38.

\bibitem[Thrun et~al., 2006]{Probabilistic-Robotics}
Thrun, S., Burgard, W., and Fox, D. (2006).
\newblock {\em {Probabilistic Robotics}}.
\newblock {The MIT Press}.

\bibitem[Urmson, 2011]{Urmson-2011}
Urmson, C. (2011).
\newblock {The Google Self-Driving Car Project}.
\newblock Technical report, {Google Inc.}

\bibitem[Vanderbilt, 2012]{agv-wired}
Vanderbilt, T. (2012).
\newblock {Let the Robot Drive The Autonomous Car of the Future is Here}.
\newblock In {\em {Wired Magazine}}.
\newblock http://www.wired.com/2012/01/ff\_autonomouscars/.

\bibitem[Vo and Ma, 2006]{GMPHD-Filter}
Vo, B. and Ma, W. (2006).
\newblock {The Gaussian Mixture Probability Hypothesis Density Filter}.
\newblock {\em {IEEE: Transactions Signal Processing}}, 54(11):4091--4104.

\end{thebibliography}

\end{document}